%% file: main_en.tex
\documentclass[fleqn,usenatbib,useAMS]{mpazh}
\usepackage{graphicx}	
\usepackage{amsmath}	
\usepackage{amssymb}	
\usepackage{multicol}        
\usepackage{bm}		

\usepackage{color}
\usepackage{pdflscape}
\usepackage[T1]{fontenc}
\usepackage{ae,aecompl}
\usepackage{newtxtext,newtxmath}
\usepackage{lscape}
\usepackage{mathtext}
\usepackage{float}
\usepackage{textcomp}
\usepackage{rotating}
\usepackage{dcolumn}
\usepackage{tabularx}
\usepackage{pdflscape}
\usepackage{url}
\usepackage{textcomp}
\usepackage{xspace}

\def\srg{SRG\xspace}
\def\srge{SRG/\textit{eROSITA}\xspace}
\def\erosita{\textit{eROSITA}\xspace}

\def\flux{erg\,s$^{-1}$\,cm$^{-2}$\xspace}

\title[Variable AGNs in SRG/eROSITA Survey]{Highly Variable Active Galactic Nuclei in the SRG/eROSITA \\ All-Sky Survey: I. Constructing a Sample and Catalog of Sources \\ Detected in Low State.}
\author[Medvedev~et~al.]{P.~S.~Medvedev$^{1}$\thanks{E-mail: \href{mailto:tomedvedev@iki.rssi.ru}{tomedvedev@iki.rssi.ru}}, 
M.~R.~Gilfanov$^{1,2}$, S.~Yu.~Sazonov$^{1}$,
\newauthor R.~A.~Sunyaev$^{1,2}$, and G.~A.~Khorunzhev$^{1}$
\\
$^{1}$ Space Research Institute of the Russian Academy of Sciences (IKI), 84/32 Profsoyuznaya Str, Moscow, Russia, 117997 \\
$^{2}$ Max Planck Institute for Astrophysics, Karl-Schwarzschild-Strasse 1, 85741 Garching, Germany}
\date{Accepted December 8, 2022. Received December 2, 2022; in original form December 2, 2022}

\pubyear{2022}

\begin{document}

\label{firstpage}
\pagerange{\pageref{firstpage}--\pageref{lastpage}}
\maketitle

\begin{abstract}
We present the results of our search for highly variable active galactic nuclei (AGNs) the X-ray flux from which changed by more than an order of magnitude during the SRG/eROSITA all-sky survey. Using the eROSITA data obtained in the period from December 2019 to February 2022, we have found 1325 sources the X-ray flux from which in the 0.3--2.3 keV energy band changed by more than a factor of 10 at a confidence level of at least 99.73\%. Of them, 635 objects have been classified as AGNs or AGN candidates. We describe the procedure of searching for highly variable sources and the selection of extragalactic objects among them and describe the statistical properties of the produced catalog. We provide a catalog of 49 sources for which a statistically significant flux in their low state was detected. For the latter we provide their light curves and X-ray spectra and discuss in detail the most interesting of them.
\end{abstract}
\begin{keywords}
 {\it supermassive black holes, accretion, active galactic nuclei, X-ray sources.}
\end{keywords}
\section{Introduction}
\label{s:intro}
One of the peculiarities of active galactic nuclei (AGNs) is significant variability of their flux at all wavelengths. In particular,  in X-rays variability of the AGN flux is observed in a wide range of characteristic time scales,  from weeks to tens of years  \citep[e.g., ][]{markowitz2004, vagnetti2011, vagnetti2016, shemmer2014}. The typical standard deviations on time scales accessible to observation with modern X-ray telescopes are $\approx 10$--20 \%, while the fractional variability amplitude rarely exceeds a factor $\sim 2$ \citep[see, e.g., ][]{gibson2012, middei2017, timlin2020}, without evidence for an explicit dependence of the variability amplitude on the AGN redshift \citep{lanzuisi2014, shemmer2017}. The X-ray variability of AGNs is known to have a red-noise power spectrum \citep[see, e.g., ][]{lawrence, mchardy, uttley}; in other words, larger variability amplitudes are observed on longer time scales. At the same time, there is evidence that the X-ray variability amplitude anticorrelates with the AGN luminosity \citep{shemmer2017}.  An analysis of the long-term X-ray variability is used to investigate the regimes of accretion disk instabilities in AGNs \citep[see, e.g., ][]{peterson2001}.

Extreme (in amplitude) changes in the X-ray flux from AGNs by a factor $\gtrsim 10$ on a time scale $\sim 1$ year are observed very rarely. Such events are well outside the overall picture of AGN variability.  The physical interpretation of such events in terms of the standard disk accretion model \citep{shakura} runs into great difficulties,  since the characteristic viscous time of an accretion disk with parameters typical for AGNs is  $\sim 10^3$ years \citep{Shakura76, Ricci_2020}.  Despite repeated reports on the detection of such abrupt changes in the X-ray flux, no systematic search for and study of such events have been carried out so far because of their rarity.

One of the possible scenarios explaining at least a (small) fraction of such events can be the transitions between various AGN states,  including  ``changing look AGN'' \citep[see, e.g., ][]{matt, Ricci_2016}.  These are the objects that change their spectral classification between type 1 and type 2 AGNs. Within the unified AGN model \citep{Antonucci},  these transitions must be caused by changes in the density of absorbing material along the line of sight blocking the broad line regions closest to the supermassive black hole (SMBH) \citep[see, e.g., ][]{Stern}.  However,  as will become clear from what follows,  this phenomenon can explain only a small fraction of the highly variable AGNs.  In this context the X-ray variability of AGNs is also of great interest, since it allows the variations of absorbing material along the line of sight to be investigated based on their X-ray spectra \citep[e.g., ][]{Puccetti_2014}.

Understanding the characteristics of the long-term X-ray variability of AGNs is also important for estimating the effects of selection on the statistical inferences made for a population of sources based on single-epoch measurements.

On July 13,  2019,  the Spectrum-Roentgen-Gamma (SRG) X-ray space observatory\citep{2021AA...656A.132S} was launched from the Baykonur launch site.  The observatory incorporates two unique mirror telescopes operating on the principle of X-ray grazing incidence: the \erosita telescope \citep{2021AA...647A...1P} operating in the soft 0.2--9.0 keV X-ray energy band and the Mikhail Pavlinsky ART-XC telescope \citep{2021AA...650A..42P} operating in the harder 4.0--30.0 keV energy band.  \erosita has a large field of view,  $\sim 1^\circ$,  and a good angular resolution,  $\sim 30$" (the diameter of the circle within which half of the photons from a point source are recorded) in the sky scanning mode. Owing to such characteristics,  it is expected that during its four-year sky survey \erosita will obtain a unique (in depth and completeness) sample of X-ray AGNs and quasars.  Owing to the strategy of the survey consisting of eight individual sky surveys with a duration of six months each,  extensive information about the variability of discovered objects will also be obtained.

This paper is devoted to searching for highly variable AGNs and quasars,  whose luminosity in the standard X-ray band changed by more than a factor of 10,  based on the results of the first two \srge all-sky surveys.  Below we will describe the procedure of searching for such sources, the selection of extragalactic objects from the produced list of highly variable sources, present our catalog, and analyze in detail those of them for which a statistically significant flux in their low state is detected.  We use the \erosita data in the eastern Galactic hemisphere $0^\circ<l<180^\circ$ the processing of which the Russian \srge consortium is responsible for.
\section{X-ray data}
\label{s:data}
We investigated the variability of the X-ray sky based on the \srge data obtained during five successive all-sky surveys conducted from December 8,  2019,  to February 20, 2022.  By this time the observatory completed four full sky surveys and surveyed approximately a third of the sky in the fifth survey ($\approx 38$\% of the full coverage).  The accumulated vignetted exposure and the achieved survey sensitivity lie in the ranges from $\sim$ 600 sec and $\sim (1-2)\cdot 10^{-14}$ \flux near the ecliptic equator and reach $\ga 10^4$ sec and $\sim (3-5)\cdot 10^{-15}$ \flux (including the confusion effect) on an area $\sim 1000$ deg$^2$ around the ecliptic poles in the 0.3--2.3 keV energy band.

The ground-based and flight calibrations obtained in the period of test observations in October--November~2019 were used to process the data. We calibrated the data and detected the sources using individual components of the eSASS software developed by the German \srge consortium \citep{brunner2022} and the software of the Russian \srge consortium developed at the Space Research Institute of the Russian Academy of Sciences. The X-ray sources were detected by fitting the distribution of counts based on the sum of all sky surveys by the maximum likelihood method using the \erosita point spread function (PSF) with the ermldet code from the eSASS software \citep{brunner2022}.

In the 0.3--2.3 keV energy band eROSITA detected more than 2 million X-ray sources with a detection likelihood $\ge6$, roughly corresponding to a significance of $\approx 3 \sigma$ for a Gaussian distribution.  The count rates recorded by the eROSITA detectors were converted to energy fluxes by assuming a power-law spectrum with a photon index $\Gamma=2.0$ and absorption $n_H=3 \times 10^{20}$ cm$^{-2}$.  The catalog of sources and the detection procedures will be described in more detail in a separate paper.  The \erosita catalog of X-ray sources constructed from the set of all sky survey data (hereafter the combined catalog) was used to search for variable sources.

In this paper we investigate the variability of the flux from sources averaged over the individual surveys.  We will restrict ourselves to the sources located at ecliptic latitudes $l_e \le 87\deg$.  At higher ecliptic latitudes the source scanning duration during one sky survey exceeds 20 days,  which is much longer than that for the sources on the ecliptic equator ($\sim 1$ day) and admits a more detailed study of their variability.  These sources will be investigated separately in our future publications.

Here we also investigate the spectral characteristics of highly variable sources.  For this purpose,  we extracted the source spectra based on data from the combined catalog for the individual \erosita surveys using an aperture with a radius of 60".  To estimate the background spectrum, we used a ring around the source with inner and outer radii of 120" and 350"--600",  respectively.  The outer radius was chosen so that there were at least 200 counts in the background region.  The sources detected in the background region were masked using a circular aperture with a radius of 30"--60", depending on their fluxes.  The direct procedure for extracting the spectra of sources was performed with the srctool code from the eSASS software. The spectra obtained were further fitted with standard tools from XSPEC \citep[v12.12,][]{Arnaud1996} using the C-statistic \citep{Cash1979} for data with a Poissonian background (the W-statistic in XSPEC).  For this purpose,  the energy channels of the spectra were first binned so that the number of recorded counts in each corresponding channel of the background spectrum was at least 5. This procedure was performed with the standard ftgrouppha tool in the HEASOFT (v6.29) software.  After the optimization of the likelihood function for the best-fit spectral model parameters, we launched Markov chain Monte Carlo (MCMC) simulations for a more accurate estimation of confidence intervals for the model parameters.  For MCMC we used the Goodman-Weare algorithm \citep{emcee}.  When launching MCMC,  we specified a chain length of $2 \cdot 10^5$ steps and 5000 steps for ``burn-in'' stage.  The convergence of the chains was checked with the Geweke convergence measure \citep{geweke}: for each model parameter this quantity was within the range from $-0.2$ to 0.2.  The uncertainties for the spectral parameters are given at a 90\% confidence level and were determined from MCMC; the best-fit parameters were determined from the likelihood maximization procedure.  The quality of the data fit by the model was estimated by the bootstrapping method with 100'000 random realizations and the Cramer-von Mises statistic \citep{cvm, anderson} using the \texttt{goodness} command in XSPEC with the ``nosim'' and ``fit'' options.

\section{Search for highly variable sources}
\label{sec:catalog}

\begin{figure}
    \centering
    \includegraphics[width=0.9\columnwidth]{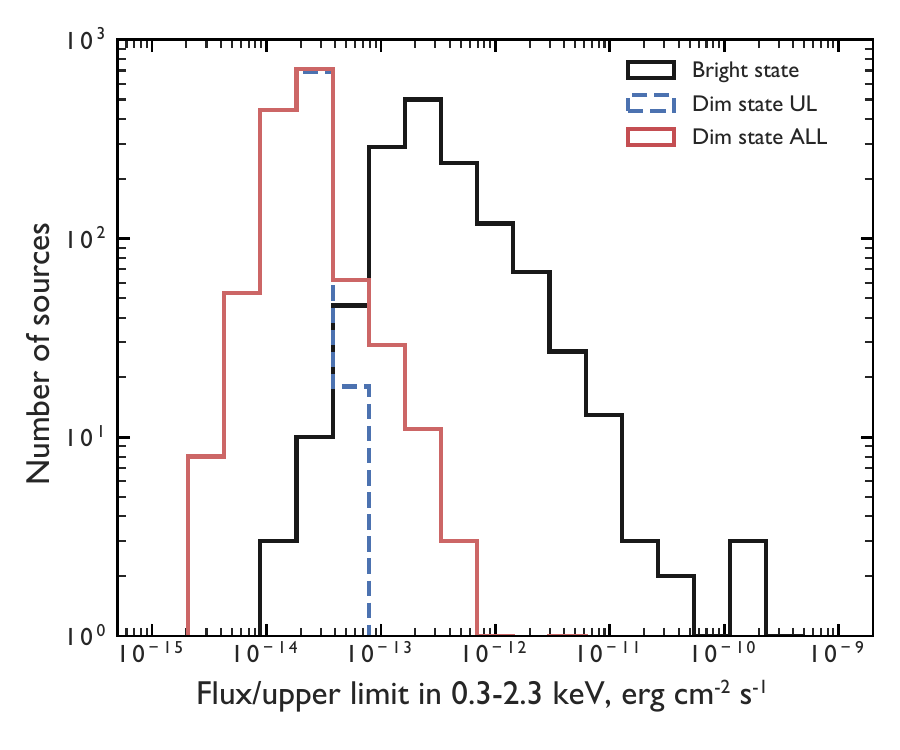}
    \begin{center}
    \caption{Flux distribution of the sample of highly variable sources in the 0.3--2.3 keV energy band.  The black histogram indicates the maximum fluxes recorded for each source among the five \srge surveys.  The red histogram indicates the minimum recorded fluxes for each source or the upper flux limits (also indicated separately by the dashed line) if the source was not detected with a likelihood exceeding the threshold value of 6 in the survey relative to which the most significant change in the flux in the bright state was obtained.}
    \label{fig:flux_distr}
   \end{center}
\end{figure}

\subsection{X-ray Light Curves}
To construct the light curves of \erosita X-ray sources,  we used the method of forced PSF photometry.  We performed forced PSF photometry based on data from each of the individual six-month sky surveys, with the positions and extents of the sources having been fixed at the values found from the sum of all surveys.  Thus,  the refitting procedure was used only to estimate the flux $F_i$ in the $i$-th sky survey and the 68\% lower and upper limits on this flux,  $F^l_i$ and $F^u_i$,  respectively.  Note that the total number of sources was fixed according to the combined catalog,  and no filtering by detection significance was performed at the stage of forced photometry.  At the same time, the lower flux limit for some sources could be zero.

\subsection{Selection Method}
To determine the confidence level of source flux change between surveys,  we assumed that the measured flux from the source was a random variable $f_i$ with a normal probability density distribution and a mean $F_i$.  The standard deviation $\sigma_i$ of the normal distribution was approximated by the mean error in the flux for the sources whose lower flux limit was above zero: $\sigma_i =  (F^u_i - F^l_i) / 2$.  Since the lower flux limit in the optimization procedure of the maximum likelihood method was set equal to zero,  the flux probability distribution for faint sources with $F_i\sim \sigma_i$ becomes asymmetric and deviates significantly from the normal distribution.  In this paper we neglected these deviations using only the upper flux limit to calculate the standard deviation for the faintest sources: $\sigma_i= F^u_i - F_i$.  This leads to some underestimation of the statistical detection significance of the flux variability relative to the surveys with low fluxes obtained by the method of forced photometry,  but does not affect significantly the selection of sources with $F_i > 3\sigma_i$,  which will mostly be discussed in this paper.  In our future publication devoted to the full catalog of highly variable AGNs, including sources with $F_i < 3\sigma_i$ in any of the surveys,  we will take into account a realistic flux probability distribution for situations with low fluxes.

Assuming that the measurements of the flux from a source in surveys $i$ and $j$ are completely independent,  the probability distribution of their ratio,  $r = f_{i} / f_{j}$,  can be written as follows:
\begin{equation}
\label{eq:p_r}
    p_r(r) = \int_{-\infty}^{+\infty} |f_j| p_{f,i}(r f_j) p_{f,j}(f_j) df_j
\end{equation}
Using the normal approximation for the flux distribution $p_{f,i}=N(F_i, \sigma_i)$ will have the following form \citep[see, e.g., ][]{hinkley1969ratio}:
\begin{multline}
\label{eq:r}
    p_r= \frac{b(r) \cdot d(r)}{a^3(r)} \frac{1}{\sqrt{2 \pi} \sigma_{i} \sigma_{j}}  \left[\Phi \left( \frac{b(r)}{a(r)}\right) - \Phi \left(-\frac{b(r)}{a(r)}\right) \right] \\ + \frac{1}{a^2(r) \cdot \pi \sigma_{i} \sigma_{j} } e^{- \frac{c}{2}},
\end{multline}
where
\begin{align*}
a(r) &= \sqrt{\frac{1}{\sigma_{i}^2} r^2 + \frac{1}{\sigma_{j}^2}} \\
b(r) &= \frac{F_{i} }{\sigma_{i}^2} r + \frac{F_{j}}{\sigma_{j}^2} \\
c &= \frac{F_{i}^2}{\sigma_{i}^2} + \frac{F_{j}^2}{\sigma_{j}^2} \\
d(r) &= e^{\frac{b^2(r) - c a^2(r)}{2a^2(r)}}
\end{align*}
and $\Phi$ is the cumulative normal distribution function
\begin{equation}
\Phi(x)= \int_{-\infty}^{x}\, \frac{1}{\sqrt{2 \pi}} e^{- \frac{1}{2} y^2} \ dy \, .
\end{equation}

For each source from the combined catalog and for each pair of surveys,  using Eq.~\ref{eq:r},  we calculated the cumulative probability that the flux ratio $r=f_i/f_j > 0.1$ provided that in the $j$-th survey the source was brighter than in the $i$-th survey,  i.e.,  $F_j>F_i$.  In other words,  we calculated the probability that the true flux from the source in the two surveys changed by less than a factor of 10.  This probability was calculated for all of the possible combinations of our five surveys.  The source passed the selection criterion and was included in our sample of highly variable sources if the probability described above was less than $2.7\cdot 10^{-3}$ at least for one pair of surveys.  Note that as the probability threshold we arbitrarily chose the value corresponding to the $3\sigma$ confidence level for a Gaussian distribution.  Taking into account the fact that the number of trials (i.e.,  the number of all possible pairs of surveys) is 10,  we expect that in the final catalog of highly variable AGNs there will be a few percent of the objects whose true variability amplitude is slightly less than 10.

\begin{figure*}
    \centering
    \includegraphics[width=0.9\textwidth]{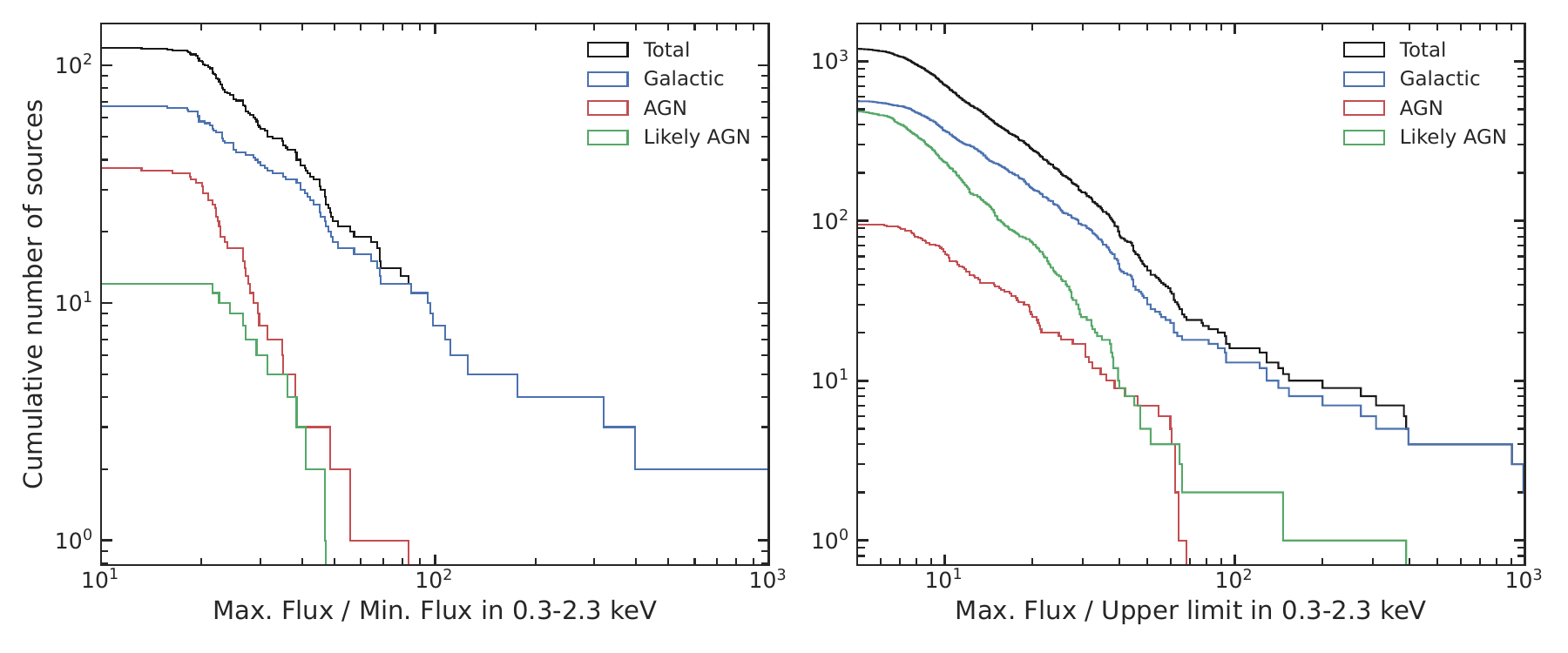}
    \begin{center}
    \caption{Distribution of the sample of highly variable sources in maximum variability amplitude $R_X$ recorded based on the five \srge surveys in the 0.3--2.3 keV energy band.  The left panel shows the sources that are detected in the surveys in which their lowest flux was recorded (likelihood $\ge 6$). In this case, $R_X$ is defined as the ratio of the maximum flux to the minimum one among all surveys.  The right panel shows the sources that are not detected in the surveys relative to which the most significant change in the flux was obtained.  For these sources $R_X$ is defined as the ratio of the maximum flux to the $3\sigma$ upper flux limit.  The black,  blue,  red,  and green lines indicate the cumulative histogram,  respectively,  for all sources,  the sources classified as Galactic ones, the spectroscopically confirmed AGNs,  and the AGN candidates selected from the remaining sources (Subsection \ref{sec:classification}).  For some of the sources shown on the right panel $R_X<10$, which is a consequence of the method of calculating the lower limit to calculate $R_X$ using the $3\sigma$ flux limit in the low state.}
    \label{fig:flux_change}
   \end{center}
\end{figure*}

\begin{table*}
\renewcommand{\arraystretch}{1.15}
\renewcommand{\tabcolsep}{0.2cm}
\linespread{1.0}
\scriptsize
\caption{The list of highly variable AGNs detected in the low state}
\label{tab:sample1}
\begin{tabular}{rlcccccc}
\hline
No. & \erosita source & $F^{max}_{0.3-2.3}$ & $R_X$ & Name &Spec. class & z & Reference   \\
 & & \flux &  & &  &   &  \\
 \hline
\input{tab_sample1}

\end{tabular}

The \erosita source name corresponds to the best X-ray coordinates; the maximum variability amplitude $R_X = F^{max}/F^{min}$ is defined as the ratio of the maximum flux to the minimum one from the source fluxes in the 0.3--2.3 keV energy band recorded among all surveys; $F^{max}_{0.3-2.2}$ is the highest X-ray flux from the source (without any correction for absorption); the spectral classes of the sources 0.3--2.2 keV and their redshifts ($z$) were determined in the papers specified in the ``Reference'' column,  for a detailed description of the notation,  see Section~\ref{sec:sample} in the text of the paper.
\end{table*}

\subsection{Sample}
The sample of highly variable sources obtained in this way contains 1325 sources.  Fig.~\ref{fig:flux_distr} shows the flux distribution for our sample of sources in the 0.3--2.3~keV energy band. The black histogram indicates the fluxes found by the method of forced photometry in the surveys in which the maximum flux was recorded (these times will be called the source bright state). The red histogram indicates the source fluxes or the upper flux limits in the surveys relative to which the most significant change in the flux was found (the source dim state). In this case, if the detection significance of a source in the dim state was below the detection likelihood threshold ($<6$), then as the source flux for the red histogram in Fig.~\ref{fig:flux_distr} we used the 3-$\sigma$ upper limit obtained from the sensitivity maps of the \erosita surveys\footnote{The sky survey sensitivity maps are calculated based on the exposure maps and background maps in the 0.3--2.3~keV energy band for the model of point sources by taking into account the point spread function and the likelihood threshold equal to 6 using the \texttt{ersensmap} task of eSASS.} (the contribution of such sources is indicated by the blue dashed histogram). Below in the paper, to parameterize the maximum variability amplitude detected from all of the sky surveys available for a given source, we will use the quantity $R_X$ defined as the ratio of the flux from the source in its bright state (black histogram) to the flux or the upper flux limit for the source in its dim state (red histogram).

\subsection{Classification of the Selected Sources}
\label{sec:classification}
Our sample of highly variable sources was cross-correlated with the Gaia~eDR3 catalog \citep{GAIAEDR3} to determine the Galactic sources. We identified an X-ray source as a Galactic one if all of the sources from the Gaia catalog falling into its position error circle had significant parallax and/or proper motion measurements with S/N$\ge 5$. When determining the confidence level of the proper motion measurement, we checked both the proper motion components in equatorial coordinates and the total proper motion of the source. As the position error circle of the X-ray sources we used a 98\% error circle; the typical radii of the position error circle for the selected \erosita sources in the bright state are $\approx 5$". As a result of this procedure, we selected 630 candidates for highly variable Galactic sources.

\begin{figure*}
    \centering
    \includegraphics[width=0.9\textwidth]{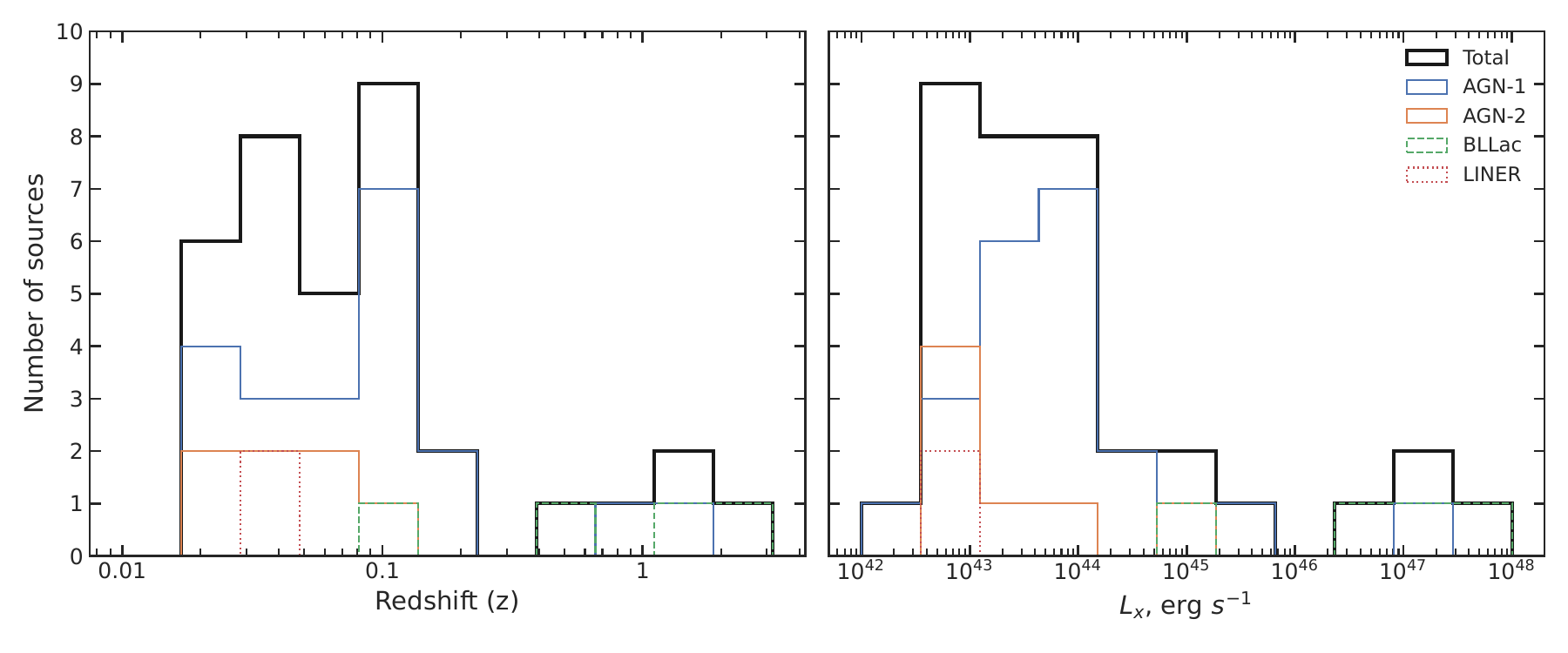}
    \begin{center}
    \caption{Left: histogram of the redshift distribution of the sources from Table~\ref{tab:sample1}. The black color indicates the combined distribution; the blue, orange, green, and red colors indicate the distributions of type 1 and 2 AGNs, BL Lac objects, and LINER galaxies, respectively. Right: the rest-frame X-ray luminosity distribution in the 0.3--10 keV energy band based on the \erosita surveys during which the maximum X-ray flux from the sources was recorded (see Table~\ref{tab:params}).}
    \label{fig:z_distr}
   \end{center}
\end{figure*}

Fig.~\ref{fig:flux_change} shows the cumulative distribution of the sources in maximum variability amplitude ($R_X$). The left panel in Fig.~\ref{fig:flux_change} shows the sources for which the detection significance in the dim state (i.e., in the survey relative to which the most significant change in the flux was recorded) was above the detection likelihood threshold ($\ge 6$). The total number of such sources was 118; we identified 67 of them as Galactic sources. In the remaining sample of 51 sources two more sources were excluded from our further analysis as candidates for tidal disruption events (TDEs) based on the spectroscopic data obtained by the \srge ground support group. 

The right panel in Fig.~\ref{fig:flux_change} shows the distribution for all of the remaining sources from our sample (1207 sources), 563 of which were identified as Galactic sources. In this case, the lower limit on the maximum variability amplitude that was calculated relative to the $3\sigma$ upper limit found from the sensitivity maps is shown. Note that for some of the sources shown on the right panel the variability amplitude $R_X<10$. This is a consequence of the method for calculating the lower limit on $R_X$ using the $3\sigma$ upper flux limit in the low state.

Our further classification of the candidates for extragalactic sources included their cross-correlation with the Million Quasars \citep[Milliquas, v7.7, ][]{milliquas1, milliquas2} catalog. The intersections of \erosita X-ray sources with sources from the Milliquas catalog and having spectroscopically confirmed classes of objects and redshifts fell into the AGN group indicated on the right and left panels in Fig.~\ref{fig:flux_change} by the red solid lines. There were a total of 132 such sources (37 and 95 on the left and right panels, respectively). From the remaining group of unidentified sources we excluded the confirmed TDEs (from the published samples by \citealt{sazonov2021} and Khorunzhev et al. and from the lists of confirmed TDEs that will be published in succeeding papers) and the known classified transient events associated with supernova explosions and GRB afterglows based on Zwicky Transient Facility (ZTF; \citealt{Bellm2019b, Graham2019, Masci2019}) and Asteroid Terrestrial-impact Last Alert System (ATLAS; \citealt{Tonry2018, Smith2020}) data. Thus, we excluded 60 more sources. The sources remaining unclassified after all of the described procedures (12 and 496 on the left and right panels, respectively) were assigned to the group of AGN candidates; they are indicated in Fig.~\ref{fig:flux_change} by the green line. Note that some of the unidentified candidates for extragalactic sources have signatures pointing to the AGN nature, such as the color $W_1-W_2>0.8$ \citep{Assef_2013} in the \textit{WISE} infrared all-sky survey \citep{Wright_2010}. Four sources from the AGN candidates on the left panel in Fig.~\ref{fig:flux_change} can be classified by this criterion as AGNs, while another 125 sources among the AGN candidates on the right panel have such a color.

In this paper we do not consider the completeness of our sample of highly variable AGNs and AGN candidates; it will be studied in more detail and described in the next paper of our series. The goal of this paper is to present a small group of the brightest and reliably selected highly variable AGNs (and AGN candidates), each of which is of scientific interest in its own right. Note, however, that the identification algorithm described above uses a 98\% radius of the position error circle for X-ray sources (the position errors of optical sources are negligible), which makes a corresponding contribution to the completeness and purity of the sample being obtained, $\sim 2$\%. Note that this contribution is not decisive. Besides, the 55 AGN candidates indicated by the green line on the right panel in Fig.~\ref{fig:flux_change} have a mixed composition of Gaia counterparts in the X-ray position error circle; at least some of them may turn out to be Galactic sources.

\begin{figure*}
\centering
\includegraphics[width=0.85\textwidth]{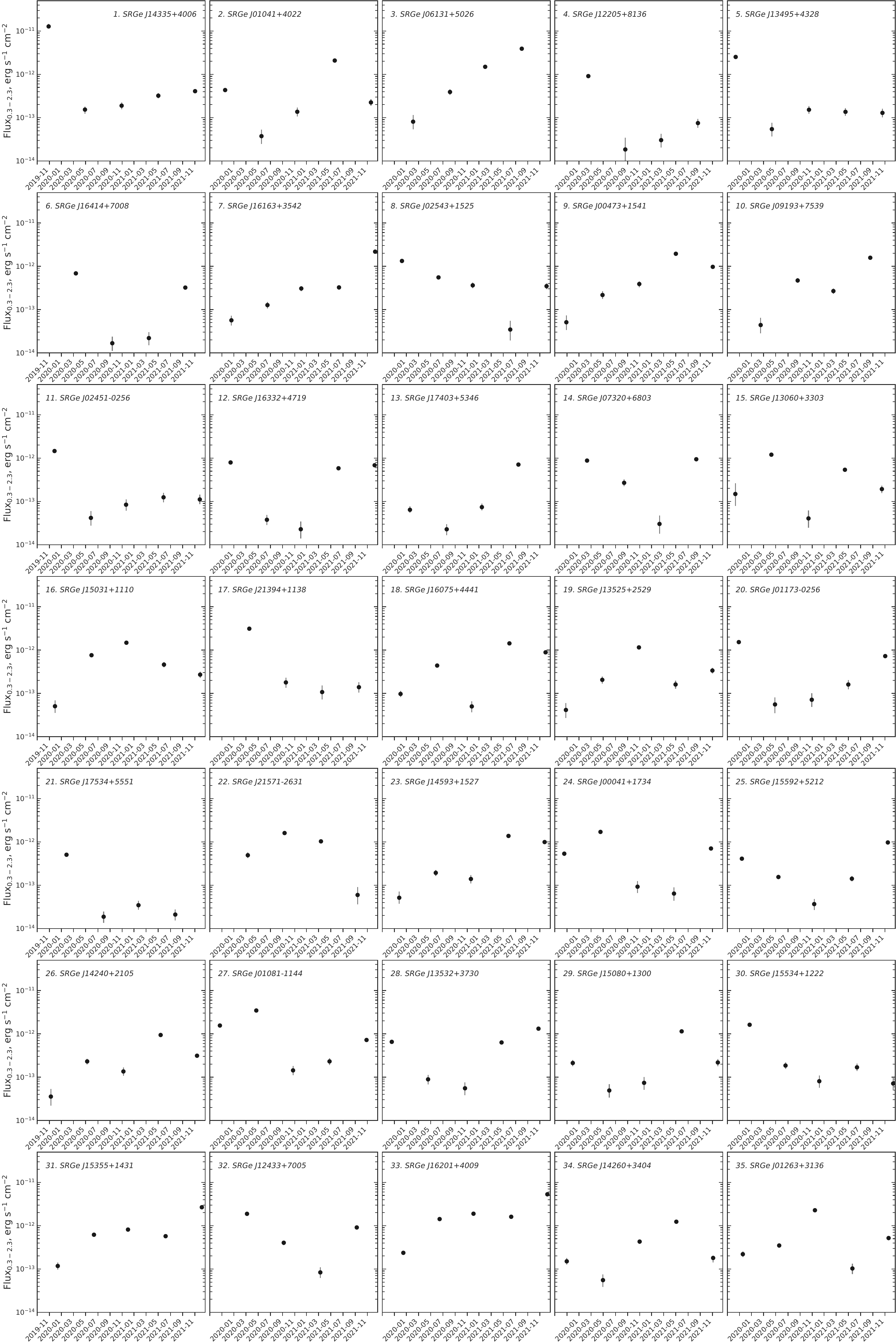}
    \caption{Long-term \srge X-ray light curves in the 0.3--2.3 keV energy band for the sample of highly variable AGNs given in Table~\ref{tab:sample1}. Each data point corresponds to the survey-averaged flux.}
    \label{fig:xraylc}
\end{figure*}

\addtocounter{figure}{-1}
\begin{figure*}
\centering
\includegraphics[width=0.85\textwidth]{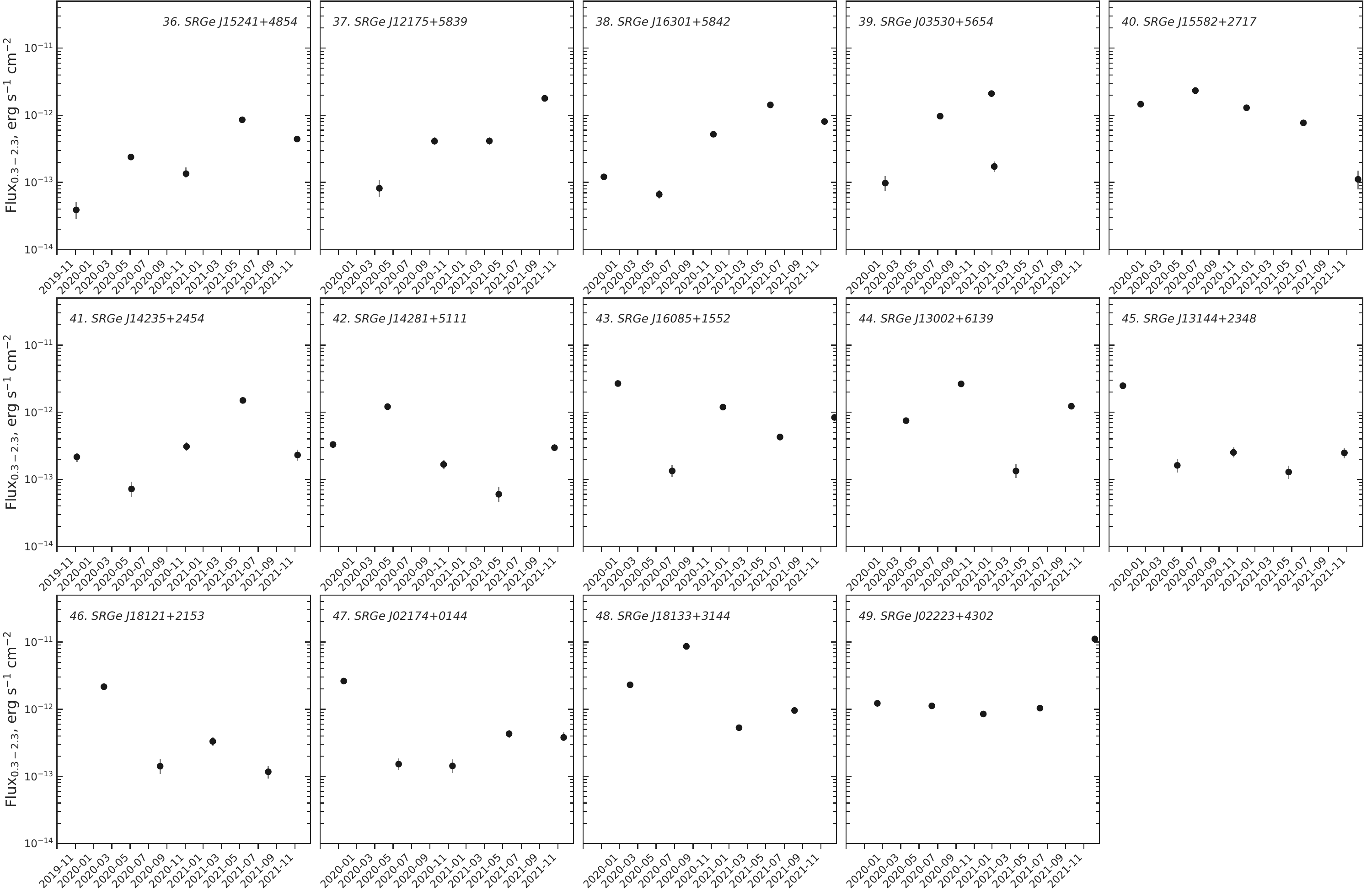}
    \caption{Continued.}
\end{figure*}

\section{Highly variable AGNs detected in their low state} 
\label{sec:sample} 
The subsequent part of the paper is devoted to investigating the subsample of highly variable AGNs and their candidates for which the low-state flux was measured with a likelihood $\ge 6$, roughly corresponding to a confidence level of $\approx 3\sigma$ for a Gaussian dis- tribution. For these objects we can calculate the variability parameter $R_X$ with a sufficient accuracy and in many cases investigate their spectral characteristics in the low state.

\subsection{Optical Classification and Redshifts}
The list of these objects and their basic parameters are given in Table~\ref{tab:sample1}. Of the 49 objects, 37 are spectroscopically confirmed AGNs, for 35 of which the redshifts were also measured, and the other 12 sources were classified by us as AGN candidates. The RX distribution of this sample has already been discussed above and corresponds to the solid red and green lines on the left panel in Fig.~\ref{fig:flux_change}. The information specified in the ``name'', ``spectral class'', and ``z'' (redshift) columns of the Table~\ref{tab:sample1} is given in accordance with the previously published papers specified in the separate ``reference'' column. The following notation is used: 
DR16 --- SDSS-DR16 \citep{dr16}, DR16Q --- SDSS-DR16Q \citep{dr16q}, LAMOST --- LAMOST Pilot Surveys \citep{lamost}, 2MAGN --- 2MASS AGN \citep{twomass}, SDSS-IV --- SDSS-IV MaNGA Sample \citep{sdss-manga}, LOZAGN --- Low-redshift AGN \citep{lozagn}, GAIA3 --- Gaia DR3 QSO candidates \citep{gaia3}, 4LAC --- Fermi AGN v4 DR3 \citep{fermi}, LAMQ5 --- LAMOST QUASAR DR5/DR4 \citep{lamost_q5}, 6dF --- 6dF galaxy survey \citep{6df}, 6dAGN --- 6dF AGN, MQ --- Milliquas \citep{milliquas1, milliquas2}, FIRST --- FIRST Bright Quasar Survey \citep{first}, Osterbrock \citep{osterbrock}, 4LAC --- Fermi AGN v4 DR3 \citep{fermi_dr3}, Veron \citep{veron}, Boisse \citep{boisse}, EXOSAT ---  EXOSAT High Galactic Latitude Survey \citep{exosat}, BZCAT --- Roma-BZCAT \citep{bzcat}.

For three sources from Table~\ref{tab:sample1} the \srge ground support group in collaboration with scientists from the California Institute of Technology obtained optical spectra at the Palomar Observatory (marked as ``Palomar'' in Table~\ref{tab:sample1}); one more source from the table was imaged with the Russian telescope at the Caucasus Mountain Observatory (CMO) of the Sternberg Astronomical Institute of the Moscow State University. The optical data on these sources will be published in a separate paper.

Among the presented sample 21 sources are spectroscopically confirmed type 1 AGNs (43\%), 7 sources are type 2 AGNs (14\%), another 7 are blazars (14\%), and 2 are LINER AGNs (4\%). The remaining 12 sources have no optical classification from the published papers to date (24\%).

\subsection{X-ray Light Curves}
Fig.~\ref{fig:xraylc} shows the \erosita X-ray light curves in the 0.3--2.3~keV energy band for the sample of sources from Table~\ref{tab:sample1}. Each light curve includes four or five flux measurements by the method of forced PSF photometry. The time interval between adjacent measurements is six months.

The light curves shown in Fig.~\ref{fig:xraylc} are characterized by a large amplitude of flux variations, in complete agreement with the source selection criterion. They exhibit a great variety in variability pattern, but their classification is complicated and, probably, will not be quite reliable and unambiguous due to the small number of measurements, 4 or 5. In some cases, astrophysically motivated classification and interpretation of the X-ray light curves can be made using spectral and (quasi-) simultaneous multiwavelength information, which our succeeding papers will be devoted to. One of the obvious questions that X-ray spectroscopy will help to answer is whether the abrupt or short-term (on time scales $\sim 6$ months) drop in the flux is associated with an increased absorption.  In particular,  in Section~\ref{sec:discussion} we will identify the class of faded sources and discuss their spectral properties.

\begin{table*}
\renewcommand{\arraystretch}{1.05}
\renewcommand{\tabcolsep}{0.2cm}
\linespread{1.0}
\begin{center}
\scriptsize
\caption{Parameters of the power-law spectral model for the sample of highly variable AGNs}
\label{tab:params}
\begin{tabular}{rlccccccc}

\hline
No. & Source & $z$ & $N_{H,gal}$ & $N_{H}$ & $\Gamma$ & $L_X$ & c-stat/d.o.f  & gof  \\
 & (SRGE...) &  & $10^{21}$ cm$^{-2}$ &  $10^{21}$ cm$^{-2}$ & & erg/s & & p-value \\
 \hline
\input{pow_fit_sample1}

\end{tabular}
\end{center}

$N_{H,gal}$ is the Galactic absorption toward the source from the HI4PI maps \citep{HI4PI_collab};
$N_{H}$ and $\Gamma$ are,  respectively,  the absorption parameter and the photon index for the best-fit power-law spectral model in the 0.3--8~keV energy band; $L_{X}$ is the rest-frame X-ray luminosity corrected for Galactic absorption in the 0.3--10~keV energy band. The characteristics of the spectra are based on the \srge survey in which the maximum flux from the source was recorded; gof (goodness-of-fit) is the probability that the observed deviation of the data from the model is the result of random fluctuations. The model with $p$-value $< 2.7 \cdot 10^{-3}$ can be rejected by the data at a $3\sigma$ confidence level. The errors correspond to a 90\% confidence level.
\end{table*}

\subsection{X-ray Spectra}
The criterion for selecting the sample of sources presented in Table~\ref{tab:sample1} allows the characteristics of their spectra to be determined not only for the surveys where the maximum flux was recorded, but also in the state with the minimum flux, although the number of such spectra is often not enough for the parameters of the spectral model to be reliably constrained. The source spectra were analyzed for each survey separately in the 0.3--8 keV energy band.

\begin{figure*}
\centering
\includegraphics[width=0.85\textwidth]{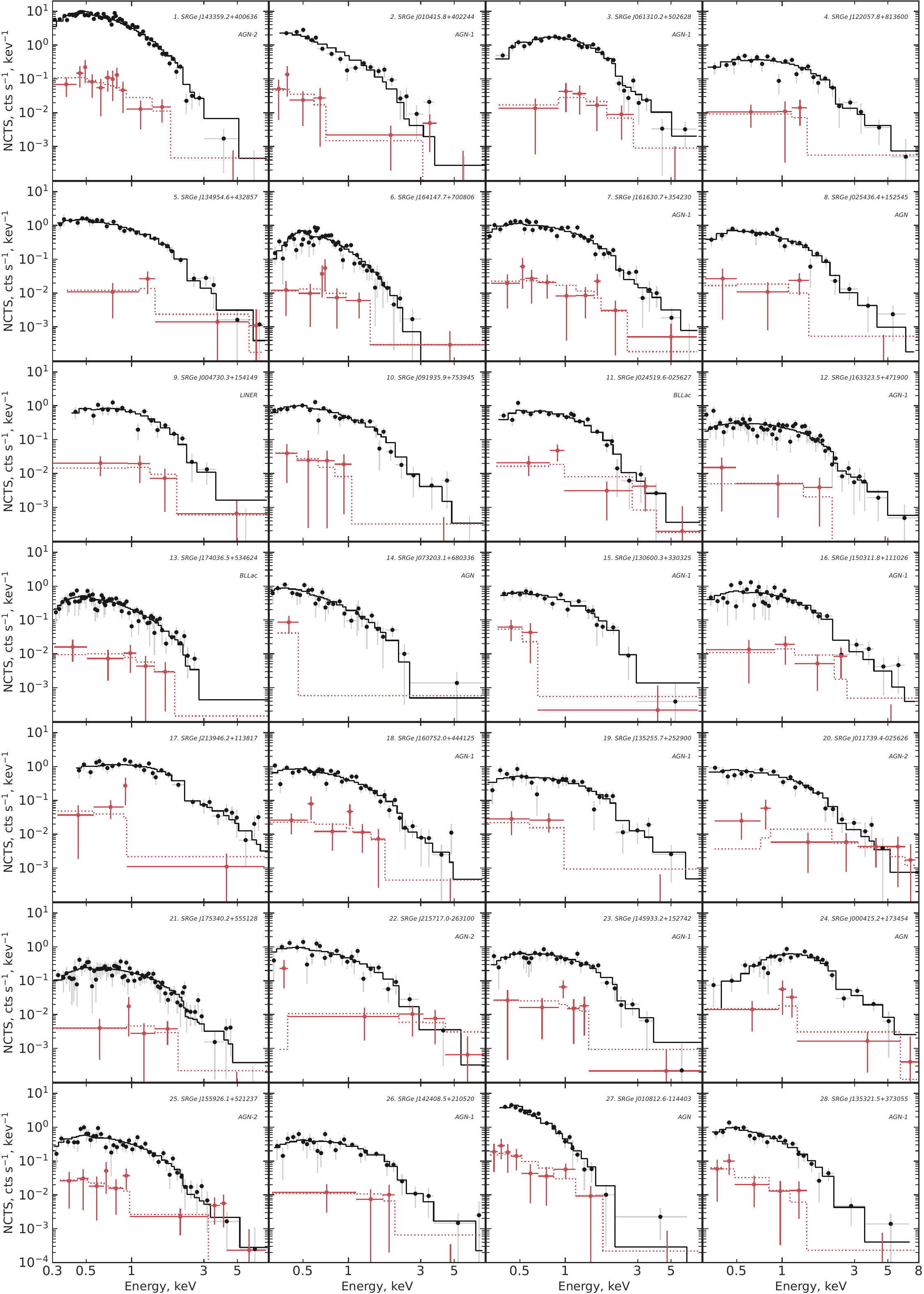}
    \caption{The X-ray spectra of objects from the sample of highly variable AGNs obtained by \erosita\ during the \srg sky surveys in which the maximum (black dots with error bars) and minimum (red dots) fluxes from the source were recorded. The solid line indicates the best-fit power-law spectral models with absorption. The parameters of the models for the high state are given in Table~\ref{tab:params}. The spectral channels were binned with a significance of at least $1\sigma$ (only for illustration).}
    \label{fig:xray_spec}
\end{figure*}

\addtocounter{figure}{-1}
\begin{figure*}
\centering
\includegraphics[width=0.85\textwidth]{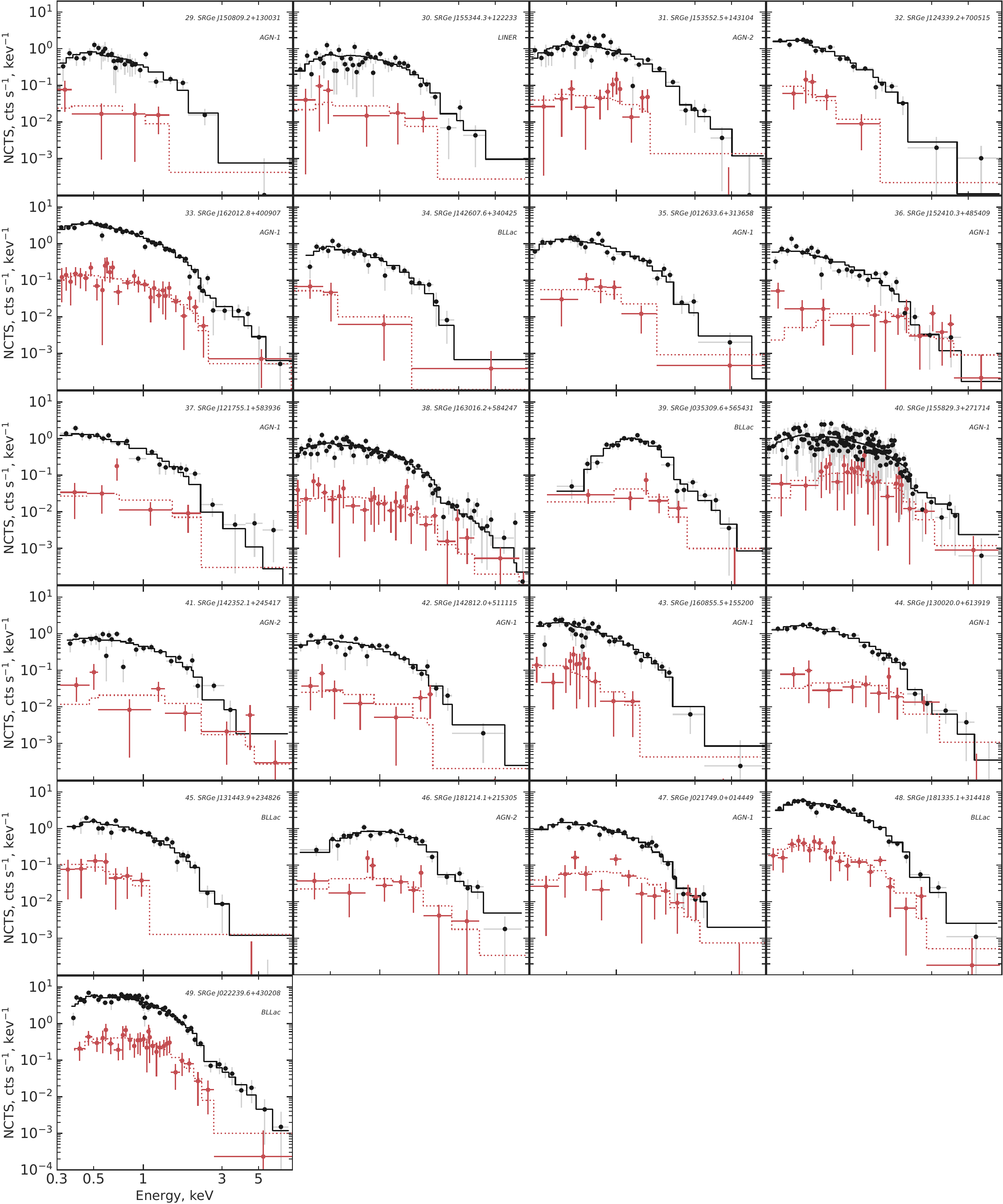}
    \caption{Continued.}
\end{figure*}

As the basic spectral model we used a power law with absorption: \texttt{tbabs*cflux*pow} in the \textsc{XSPEC} notation. The absorption parameter $N_{\rm H}$ was determined when fitting the spectral model for the spectra with at least 50 counts in the energy band under consideration; in other cases,  the parameter was fixed at the values equal to the Galactic absorption toward the sources based on the HI4PI maps \citep{HI4PI_collab}.  The elemental abundances in the absorption model were specified in accordance with \cite{wilms2000}.  The parameters for our best-fit models for the bright state of the sources are given in Table~\ref{tab:params}.  In the bright state for all sources $N_{\rm H}$ was determined from the data (all spectra have $>50$ counts). The spectra of the sources in the bright and dim states,  along with the corresponding best-fit power-law models, are shown in Fig.~\ref{fig:xray_spec}.  Note that the spectra of some sources in the bright state are poorly described by the power-law model.  In particular,  as follows from the derived goodness-of-fit values,  the power-law model with absorption can be rejected at a confidence level $>3\sigma$ for the sources  SRGE\,J1433+4006 (source no. 1 in Table~\ref{tab:params}), SRGE\,J0104+4022 (no. 2), and SRGE\,J1641+7008 (no.6).  The goal of using the power-law model here is a homogeneous characterization of the spectral hardness of the objects from the presented sample.  Note also that within the power-law model the absorption parameter in many cases exceeds noticeably the expected Galactic absorption,  at a confidence level $\ge 90\%$ for 15 sources.

Using the fitting results for the \texttt{cflux} model component, we calculated the rest-frame X-ray luminosities of the sources in the 0.3--10 keV energy band corrected for the absorption in the Galactic interstellar medium.  The following cosmological parameters were used to calculate the luminosities: $H_0=70$ km\,s$^{-1}$ and $\Omega_M = 0.3$. Table~\ref{tab:params} gives the luminosities only for the sources with spectroscopically measured redshifts.  Fig.~\ref{fig:z_distr} shows the distribution of the sample of sources from Table~\ref{tab:sample1} in redshift (left panel) and X-ray luminosity (right panel) for the bright states.  The redshifts and X-ray luminosities of the presented sample of sources lie in the ranges $z=0.018$--$2.06$ and $L_X = 3\cdot10^{42}$--$3\cdot10^{47}$ erg\,s$^{-1}$,  respectively.

\begin{figure}
\centering
\includegraphics[width=0.9\columnwidth]{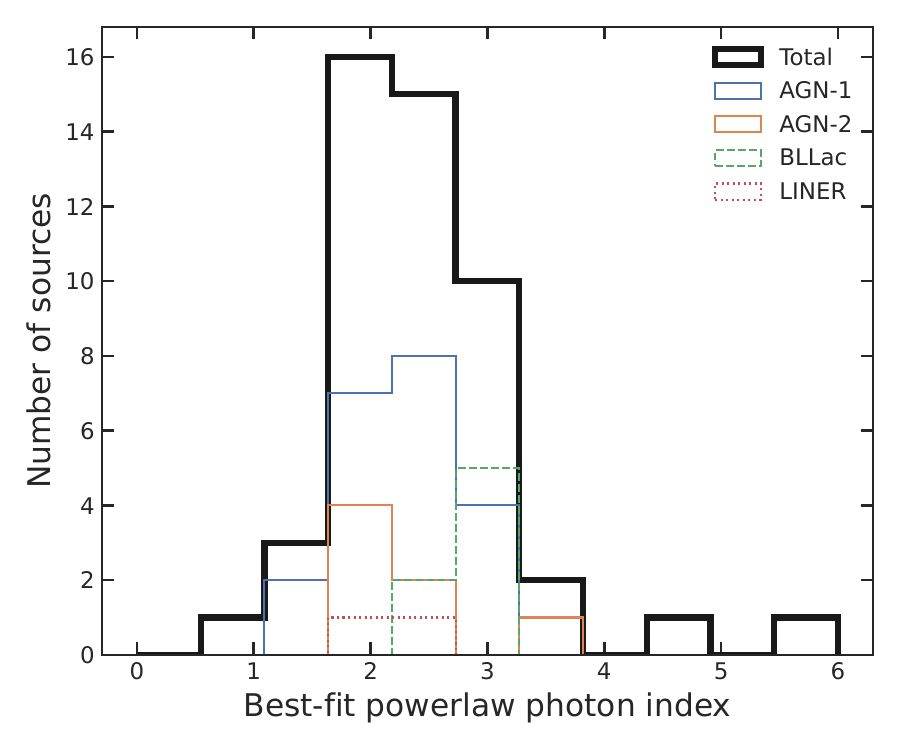}
\begin{center}
\caption{Distribution of the sources in photon index for the power-law spectral model with absorption in the states with the maximum flux (see Table~\ref{tab:params}).  The black color indicates the combined distribution; the blue, orange,  green, and red colors indicate the distributions of type 1 and 2 AGNs,  BL Lac objects,  and LINER galaxies,  respectively (see Table~\ref{tab:sample1}).}
\label{fig:g_distr}
\end{center}
\end{figure}

\begin{figure}
\centering
\includegraphics[width=0.85\columnwidth]{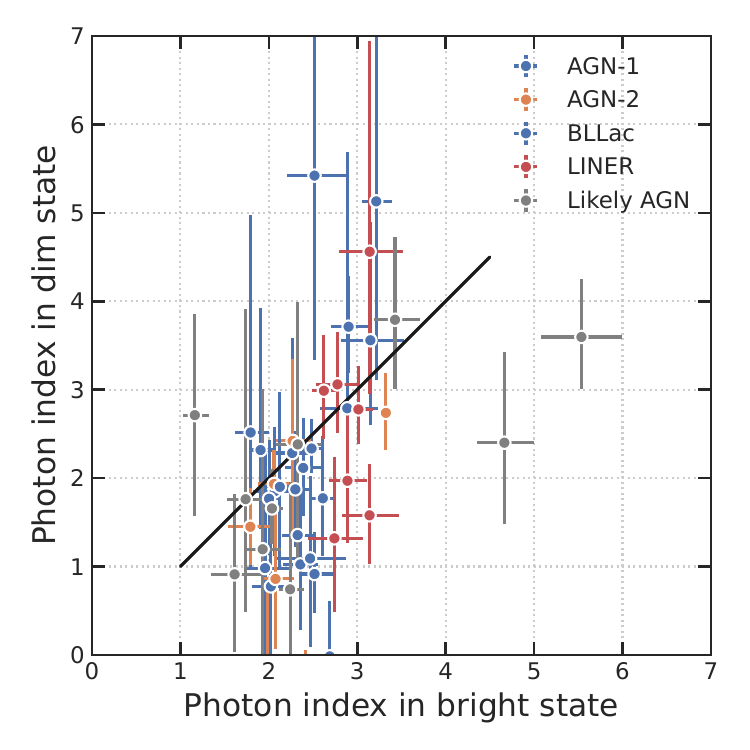}
\begin{center}
\caption{Comparison of the photon indices of the power-law model for the objects from Table~\ref{tab:sample1} in the states with the minimum (vertical axis) and maximum (horizontal axis) fluxes. The $1\sigma$ errors.  Only the objects for which the sum of the upper and lower measurement errors of the photon index in the low state does not exceed 4 are shown.}
\label{fig:gmin_gmax}
\end{center}
\end{figure}

Fig.~\ref{fig:g_distr} shows the distribution of the sources in photon index for the bright-state spectra. The median of the distribution corresponds to a photon index of 2.3,  while 68\% of the sample have photon indices in the range from 2 to 3.  We can note three sources that lie well outside this distribution: SRGE\,J1641+7008 (no. 6) and SRGE\,J0108-1144 (no. 27) with photon indices of $4.63_{-0.46}^{+0.60}$ and $5.48_{-0.68}^{+0.83}$,  respectively,  SRGE\,J2139+1138 (no. 17) with a photon index of $1.2\pm0.1$.  Below we will consider their spectra in more detail.

\begin{figure*}
\centering
\includegraphics[width=0.63\textwidth]{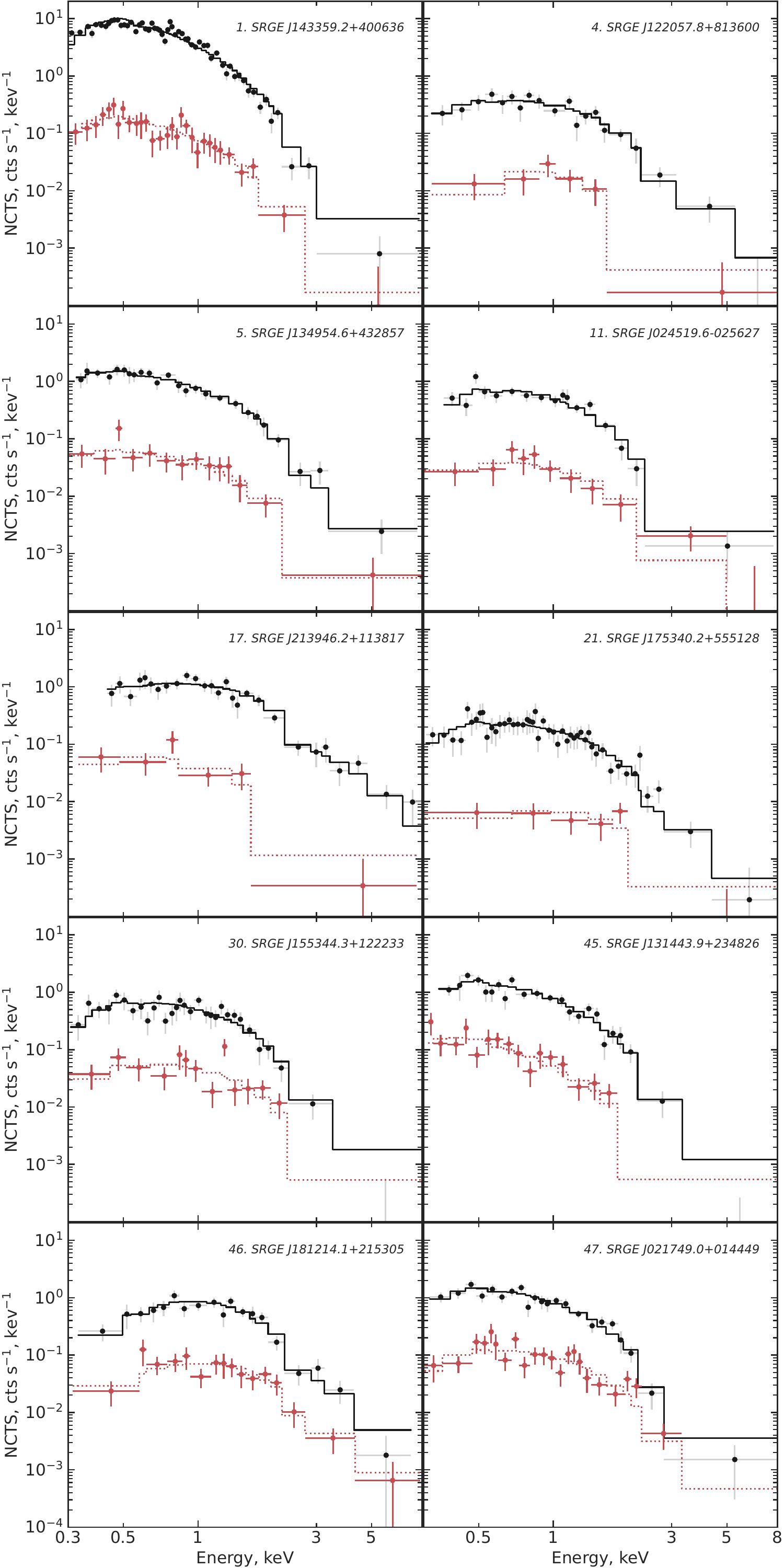}
    \caption{The X-ray spectra of the faded AGNs in the second survey.  The black color indicates the high-state spectra (the first survey); the red color indicates the combined low-state spectra (surveys 2--5).  The spectral channels were binned with a significance of at least $2\sigma$ (only for illustration).}
    \label{fig:faders}
\end{figure*}

Fig.~\ref{fig:gmin_gmax} compares the derived photon indices of the sources during their bright and dim states. An asymmetry in the distribution of data points is clearly seen: the high-state spectra are, on average, softer than the low-state ones. Indeed, comparing the a posterior probability distribution for the parameter $\Gamma$ from the spectrum during the survey with the maximum flux from the source with the distribution from the survey with the minimum flux,  we detected nine sources for which an increase in the photon index in the bright state is observed at a confidence level $>90\%$, namely sources no. 3, 6, 20, 22, 27, 36, 39, 44, and 47.  The most significant ($>3\sigma$) change in the photon index was found for the sources SRGE\,J1524+4854 (no. 36) and SRGE\,J1300+6139 (no. 44).  The following low-state model parameters were obtained for them: $\Gamma=-0.05_{-0.91}^{+1.11}$ and $\Gamma=0.87_{-0.67}^{+0.84}$.  For the same confidence level (90\%) we detected no source from our sample for which a statistically significant decrease in the photon index in the bright state would be observed (but two such cases were detected among the faded AGNs when adding the data of several surveys in the low state,  see the next section).

\section{Discussion of individual sources}
\label{sec:discussion}
In this section we discuss the most interesting sources from the catalog in Table~\ref{tab:sample1}.

\subsection{Fading AGNs}
\begin{figure}
\centering
\includegraphics[width=0.85\columnwidth]{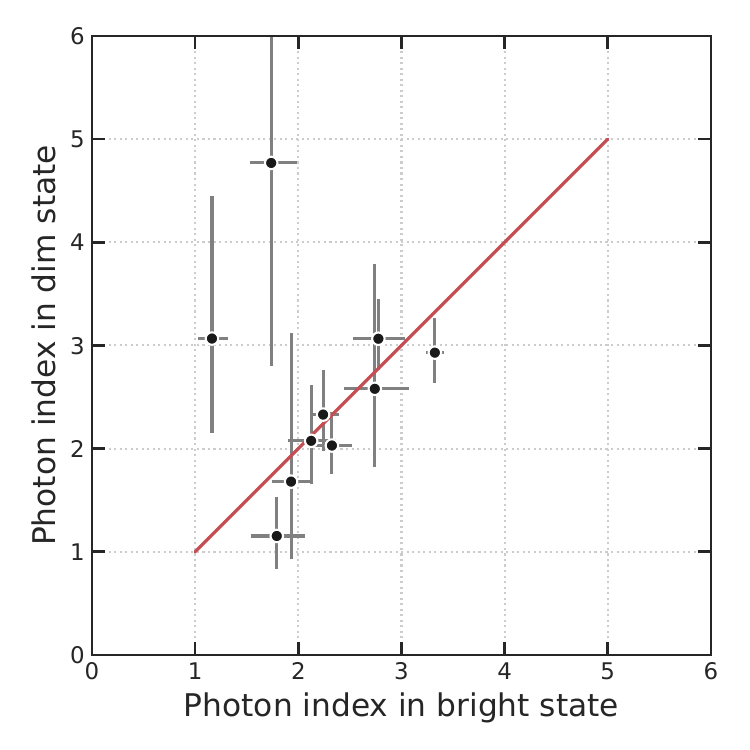}
\begin{center}
\caption{Comparison of the photon indices of the power-law model for the sample of faded AGNs.  The photon index for the spectra of the sources in the first survey (bright state) and for the spectra from the sum of surveys 2--5 (dim state) is shown along the horizontal and vertical axes, respectively.  The errors are given at a 68\% confidence level.}
\label{fig:gmin_gmax_faders}
\end{center}
\end{figure}

Based on the pattern of the light curves shown in Fig.~\ref{fig:xraylc},  among them we can identify the group of faded sources for which a significant drop in the X-ray flux was recorded between two surveys,  whereupon in the succeeding surveys the flux was not restored to values comparable to the maximum one.  Such variability can be the result of a source outburst or,  alternatively,  a longer transition from the high to low state. As has already been noted above, this separation is arbitrary,  since,  interpreting 4 or 5 flux measurements, we cannot rule out a more complex picture of variability on long time scales.  Nevertheless,  it is of interest to compare the X-ray spectra of such sources in the high and low states.  For this purpose,  without claiming to be complete,  we identified ten sources using the following selection criteria: (i) the source flux decreased in the second survey relative to the first one by more than a factor of 8 and (ii) in all of the subsequent observations it remained no more than 20\% of the flux in the first survey.  To improve the statistical quality of the low- state spectra,  we combined the data from all of the surveys in which the source was in the low state.  The high- and low-state spectra obtained in this way are presented in Fig.~\ref{fig:faders},  while the best-fit photon indices in the high and low states are presented in Fig.~\ref{fig:gmin_gmax_faders}.

As follows from Fig.~\ref{fig:faders}, the dramatic drops in the source luminosity,  as a rule,  are not accompanied by significant spectral changes. Thus,  we immediately rule out the great absorption variations as a cause of the flux changes (in the sources under consideration).  Two sources exhibited a change in the spectral shape of moderate statistical significance: SRGE\,J1220+8136 (no. 4) and SRGE\,J2139+1138 (no. 17, Fig.~\ref{fig:gmin_gmax_faders}).  Interestingly, these two sources are characterized by the hardest high-state spectra (see Table~\ref{tab:params}).  In this sense they show a trend opposite to most of the sources under consideration (Fig.~\ref{fig:gmin_gmax}).  We found an increase in the photon index in the power-law model with absorption in the low state for these two sources and an increase in the absorption column density $N_H$ at a confidence level $>90$\% for the source SRGE\,J122057.8+813600 (according to the probability distributions of the parameters found from MCMC).  The parameters of the power-law models in the low state for them are: $\Gamma = 3.4_{-0.6}^{+4.0}$, $N_{H}=4.4_{-0.9}^{+8.8}\cdot10^{21}$ cm$^{-2}$ for SRGE\,J122057.8+813600 and $\Gamma=2.4_{-0.2}^{+2.1}$, $N_{H}=1.0_{-0.2}^{+3.7}\cdot10^{21}$ cm$^{-2}$ for SRGE\,J213946.2+113817 (the error corresponds to a 90\% confidence level).  For other faded sources we failed to detect statistically significant changes in their spectral parameters (see Fig.~\ref{fig:gmin_gmax_faders}).

\subsection{Sources with Anomalously Soft Spectra}
Other interesting objects in Table~\ref{tab:sample1} are the sources SRGE\,J1641+7008 (no. 6) and SRGE\,J0108-1144 (no. 27) having the softest high-state spectra with photon indices $\Gamma\approx 4.6\pm 0.3$ and  $\Gamma\approx 5.7\pm 0.5$, respectively (see Table~\ref{tab:params}).  We fitted the high-state spectra of these sources by the simplest accretion disk model (\textsc{diskbb} in XSPEC, more complex models are more difficult to apply due to the unknown redshifts of these sources) and obtained the following parameters:
$T_{in} = 0.21_{-0.03}^{+0.02}$, $n_{H} = 0.68_{-0.28}^{+0.53} \times 10^{21}$ cm$^{-2}$ for SRGE\,J164147.7+700806 and $T_{in} = 0.15_{-0.02}^{+0.01}$, $n_{H} < 0.68 \times 10^{21}$ cm$^{-2}$ for SRGE\,J010812.6-114403.  Note that for the first source the thermal model describes the observed spectra significantly better: $\delta$c-stat$ = -7.0$.  For the second source no improvement in the quality of the fit was found: $\delta$c-stat$ = -0.1$.

Both sources have no published spectral classifications and redshifts. The X-ray position of
SRGE\,J1641+7008 is compatible with the galaxy LEDA\,2734260,  which is also the known radio source VLAJ164147.34+700806.1.

The softness of the high-state spectra could suggest that they are associated with TDEs (Sazonov et al. 2021).  However, the observed light curves of these sources are atypical for TDEs.  Note also that we detected no optical transients for these sources based on open ZTF data.  These sources require a further study,  in particular,  in the optical range.

\subsection{SRGE\,J035309.6+565431}
The source SRGE\,J0353+5654 (no. 39) possesses the highest absorption among the objects of our sample based on the power-law model fit (see Table~\ref{tab:params}).  Its X-ray position is compatible with the known blazar GB6\,J0353+5654, which belongs to the class of blazars with a high-frequency synchrotron peak (HSP).  Its optical spectrum is featureless,  the redshift is unknown \citep{crespo_2016}.  No optical transients were detected in the position error circle either.

The X-ray light curve of SRGE\,J0353+5654 (see Fig.~\ref{fig:xraylc}) exhibits abrupt changes in the flux in all four surveys. The simple thermal models with absorption describe the spectrum slightly more poorly than does the absorbed power-law spectrum,  $\delta$c-stat$ = +11.6$ (\textsc{bbodyrad}) and  $\delta$c-stat$ = +5.6$ (\textsc{diskbb}),  with the absorption parameter for these models being consistent with the Galactic value toward the source.  We obtained the following parameters for the \textsc{diskbb} model: $T_{in}=0.67_{-0.13}^{+0.11}$ keV and $N_H=6.5_{-1.0}^{+1.7} \times 10^{21}$ cm$^{-2}$.  However,  the physical motivation for the application of thermal spectral models to the description of the spectra for blazars is not obvious.

\section{Conclusions}
Although the X-ray flux variability is an integral property of accretion and,  in particular,  accretion onto SMBHs in AGNs,  dramatic,  by more than an order of magnitude, changes in the AGN luminosity on a time scale $\sim1$ year are very rare.  Moreover,  in the standard picture of accretion onto a SMBH \citep{shakura} their existence is difficult to explain.  In view of their rarity,  such events have been poorly studied, in particular,  there is no estimate of their frequency and there is no clear picture of their abundance among AGNs of various types.  The \srge all-sky X-ray survey for the first time has provided an opportunity to systematically study such events,  to estimate the frequency of their occurrence and the association with specific AGN classes.

This paper is the first in our series of works aimed at searching for and studying highly variable AGNs and quasars based on the \srge all-sky survey data. We described the method of searching for such objects in the \erosita catalog of X-ray sources and the methods of selecting AGNs and their candidates and analyzed the primary statistical properties of our preliminary catalog of 1365 highly variable objects.

Of the 1365 highly variable objects, for 630 we confirmed their Galactic origin with sufficient reliability,  for another 60 we confirmed their association with tidal disruption events, and 675 are AGNs or their candidates.  Thus,  taking into account the fact that more than a million sources in the \erosita catalog are AGNs and quasars, the fraction of such highly variable objects among AGNs is less than $\la 7\cdot 10^{-4}$,  confirming the above assertion about the extreme rarity of such events.

In this paper we presented our catalog and studied in detail 49 highly variable AGNs and their candidates for which a statistically significant flux is detected in their low state. Many of these objects have been well studied in the optical range. Our analysis of their \erosita X-ray spectra and the optical classification at this stage has not revealed obvious trends and predominant associations with specific AGN types.  However, we can conclude already now that they are not associated predominantly with blazars. It is also interesting to note that the high states of AGNs,  on average,  are characterized by softer X-ray spectra.

Our study of these unique objects will be continued in the succeeding papers of this series.

\section*{Acknowledgements}
This work is based on observations with the \erosita\ telescope on board the \srg\ observatory. The \srg\ observatory was built by Roskosmos in the interests of the Russian Academy of Sciences represented by its Space Research Institute (IKI) in the framework of the Russian Federal Space Program, with the participation of the Deutsches Zentrum f\"{u}r Luft- und Raumfahrt (DLR). The \srg/\erosita\ X-ray telescope was built by a consortium of German Institutes led by MPE, and supported by DLR. The \srg\ spacecraft was designed, built, launched and is operated by the Lavochkin Association and its subcontractors. The science data are downlinked via the Deep Space Network Antennae in Bear Lakes, Ussurijsk, and Baykonur, funded by Roskosmos. The \erosita\ data used in this work were processed using the eSASS software system developed by the German \erosita\ consortium and proprietary data reduction and analysis software developed by the Russian \erosita\ Consortium. This work was supported by RSF no. 21-12-00343.

\bibliographystyle{mnras}
\bibliography{pazh_bib_utf8} 

\end{document}

%% file: tab_sample1.tex
1 & SRGE\,J143359.2+400636 & $ 1.3 \times 10^{-11} $ & 83.3 & PGC 2160796 & AGN-2 & 0.099 & DR16 \\[0.1cm]
2 & SRGE\,J010415.8+402244 & $ 2.1 \times 10^{-12} $ & 55.6 & LAMOSTJ010415.77+402243.9 & AGN-1 & 0.193 & LAMOST \\[0.1cm]
3 & SRGE\,J061310.2+502628 & $ 3.9 \times 10^{-12} $ & 48.4 & NPM 1G+50.0043 & AGN-1 & 0.019 & 2MAGN \\[0.1cm]
4 & SRGE\,J122057.8+813600 & $ 9.1 \times 10^{-13} $ & 47.2 &  &  &  &  \\[0.1cm]
5 & SRGE\,J134954.6+432857 & $ 2.5 \times 10^{-12} $ & 46.4 & LEDA 2221805 &  & 0.045 & SDSS-IV \\[0.1cm]
6 & SRGE\,J164147.7+700806 & $ 6.8 \times 10^{-13} $ & 41.0 & NPM 1G+70.0164 &  &  & MQ \\[0.1cm]
7 & SRGE\,J161630.7+354230 & $ 2.2 \times 10^{-12} $ & 38.4 & NGC 6104 & AGN-1 & 0.028 & LOZAGN \\[0.1cm]
8 & SRGE\,J025436.4+152545 & $ 1.3 \times 10^{-12} $ & 38.4 & PGC 1484380 &  &  & CMO \\[0.1cm]
9 & SRGE\,J004730.3+154149 & $ 1.9 \times 10^{-12} $ & 38.1 & MCG+02-03-002 & LINER & 0.031 & DR16 \\[0.1cm]
10 & SRGE\,J091935.9+753945 & $ 1.6 \times 10^{-12} $ & 35.8 & GAIA 1124646674042078464 &  & (0.5) & GAIA3 \\[0.1cm]
11 & SRGE\,J024519.6-025627 & $ 1.5 \times 10^{-12} $ & 35.0 & SDSS J024519.63-025628.1 & BLLac &  & DR16 \\[0.1cm]
12 & SRGE\,J163323.5+471900 & $ 7.9 \times 10^{-13} $ & 34.8 & IRAS 16319+4725 & AGN-1 & 0.116 & LOZAGN \\[0.1cm]
13 & SRGE\,J174036.5+534624 & $ 7.1 \times 10^{-13} $ & 31.2 & 87GB 173932.3+534742 & BLLac &  & 4LAC \\[0.1cm]
14 & SRGE\,J073203.1+680336 & $ 9.4 \times 10^{-13} $ & 31.2 &  &  &  & Palomar \\[0.1cm]
15 & SRGE\,J130600.3+330325 & $ 1.2 \times 10^{-12} $ & 29.6 & PGC 2022001 & AGN-1 & 0.088 & DR16Q \\[0.1cm]
16 & SRGE\,J150311.8+111026 & $ 1.5 \times 10^{-12} $ & 29.3 & SDSS J150311.67+111027.3 & AGN-1 & 0.043 & LOZAGN \\[0.1cm]
17 & SRGE\,J213946.2+113817 & $ 3.1 \times 10^{-12} $ & 29.1 &  &  &  &  \\[0.1cm]
18 & SRGE\,J160752.0+444125 & $ 1.4 \times 10^{-12} $ & 28.6 & LAMOSTJ160751.95+444124.6 & AGN-1 & 0.718 & LAMQ5 \\[0.1cm]
19 & SRGE\,J135255.7+252900 & $ 1.1 \times 10^{-12} $ & 27.8 & KUG 1350+257 & AGN-1 & 0.064 & LOZAGN \\[0.1cm]
20 & SRGE\,J011739.4-025626 & $ 1.5 \times 10^{-12} $ & 27.5 & GJ0117394-025627 & AGN-2 & 0.051 & 6dAGN \\[0.1cm]
21 & SRGE\,J175340.2+555128 & $ 5.1 \times 10^{-13} $ & 27.0 & LEDA 2516819 &  &  &  \\[0.1cm]
22 & SRGE\,J215717.0-263100 & $ 1.6 \times 10^{-12} $ & 26.9 & 2MASX J21571697-2630596 & AGN-2 & 0.034 & 6dF \\[0.1cm]
23 & SRGE\,J145933.2+152742 & $ 1.4 \times 10^{-12} $ & 26.7 & PGC 1485268 & AGN-1 & 0.074 & LOZAGN \\[0.1cm]
24 & SRGE\,J000415.2+173454 & $ 1.7 \times 10^{-12} $ & 26.5 &  &  &  & Palomar \\[0.1cm]
25 & SRGE\,J155926.1+521237 & $ 9.7 \times 10^{-13} $ & 26.5 & SDSS J155926.11+521235.2 & AGN-2 & 0.042 & LOZAGN \\[0.1cm]
26 & SRGE\,J142408.5+210520 & $ 9.4 \times 10^{-13} $ & 26.4 & PGC 1642489 & AGN-1 & 0.047 & LOZAGN \\[0.1cm]
27 & SRGE\,J010812.6-114403 & $ 3.4 \times 10^{-12} $ & 24.1 & IRAS 01056-12009 &  &  & Palomar \\[0.1cm]
28 & SRGE\,J135321.5+373055 & $ 1.3 \times 10^{-12} $ & 23.8 & PGC 2102603 & AGN-1 & 0.107 & LOZAGN \\[0.1cm]
29 & SRGE\,J150809.2+130031 & $ 1.1 \times 10^{-12} $ & 23.2 & SDSS J150809.20+130032.4 & AGN-1 & 0.086 & LOZAGN \\[0.1cm]
30 & SRGE\,J155344.3+122233 & $ 1.6 \times 10^{-12} $ & 22.7 & 2MASX J15534435+1222337 & LINER & 0.035 & DR16 \\[0.1cm]
31 & SRGE\,J153552.5+143104 & $ 2.7 \times 10^{-12} $ & 22.7 & AKN 479 & AGN-2 & 0.020 & LOZAGN \\[0.1cm]
32 & SRGE\,J124339.2+700515 & $ 1.9 \times 10^{-12} $ & 22.5 & WISEA J124339.43+700517.0 &  &  & MQ \\[0.1cm]
33 & SRGE\,J162012.8+400907 & $ 5.3 \times 10^{-12} $ & 22.4 & KUG 1618+402 & AGN-1 & 0.028 & LOZAGN \\[0.1cm]
34 & SRGE\,J142607.6+340425 & $ 1.2 \times 10^{-12} $ & 22.3 & CSO 450 & BLLac & 1.553 & FIRST \\[0.1cm]
35 & SRGE\,J012633.6+313658 & $ 2.3 \times 10^{-12} $ & 22.1 & MCG 5-04-059 & AGN-1 & 0.045 & Osterbrock \\[0.1cm]
36 & SRGE\,J152410.3+485409 & $ 8.6 \times 10^{-13} $ & 22.0 & SDSS J152410.34+485409.7 & AGN-1 & 0.145 & LOZAGN \\[0.1cm]
37 & SRGE\,J121755.1+583936 & $ 1.8 \times 10^{-12} $ & 21.8 & CGCG 293-9 & AGN-1 & 0.023 & LOZAGN \\[0.1cm]
38 & SRGE\,J163016.2+584247 & $ 1.4 \times 10^{-12} $ & 21.5 & NPM 1G+58.0189 &  &  &  \\[0.1cm]
39 & SRGE\,J035309.6+565431 & $ 2.1 \times 10^{-12} $ & 21.5 & GB6 J0353+5654 & BLLac &  & 4LAC \\[0.1cm]
40 & SRGE\,J155829.3+271714 & $ 2.3 \times 10^{-12} $ & 20.9 & PGC 1803429 & AGN-1 & 0.090 & LOZAGN \\[0.1cm]
41 & SRGE\,J142352.1+245417 & $ 1.5 \times 10^{-12} $ & 20.8 & SDSS J142352.08+245417.1 & AGN-2 & 0.074 & DR16 \\[0.1cm]
42 & SRGE\,J142812.0+511115 & $ 1.2 \times 10^{-12} $ & 20.2 & SDSS J142811.88+511116.6 & AGN-1 & 0.129 & LOZAGN \\[0.1cm]
43 & SRGE\,J160855.5+155200 & $ 2.7 \times 10^{-12} $ & 20.0 & SDSS J160855.60+155200.2 & AGN-1 & 0.115 & DR16Q \\[0.1cm]
44 & SRGE\,J130020.0+613919 & $ 2.6 \times 10^{-12} $ & 19.8 & MCG 10-19-011 & AGN-1 & 0.052 & LOZAGN \\[0.1cm]
45 & SRGE\,J131443.9+234826 & $ 2.5 \times 10^{-12} $ & 19.3 & TXS 1312+240 & BLLac & 2.060 & DR16Q \\[0.1cm]
46 & SRGE\,J181214.1+215305 & $ 2.2 \times 10^{-12} $ & 18.5 & CGCG 142-19 & AGN-2 & 0.018 & Veron \\[0.1cm]
47 & SRGE\,J021749.0+014449 & $ 2.6 \times 10^{-12} $ & 18.4 & PKS 0215+015 & AGN-1 & 1.715 & Boisse \\[0.1cm]
48 & SRGE\,J181335.1+314418 & $ 8.6 \times 10^{-12} $ & 16.3 & B2 1811+31 & BLLac & 0.117 & EXOSAT \\[0.1cm]
49 & SRGE\,J022239.6+430208 & $ 1.1 \times 10^{-11} $ & 13.1 & 3C 66A & BLLac & 0.444 & BZCAT \\[0.1cm]

%% file: pow_fit_sample1.tex
1 & J143359.2+400636 & 0.099 & 0.10 & $ 0.85_{-0.14}^{+0.15} $ & $ 3.32_{-0.13}^{+0.14} $ & $ 8.09_{-0.79}^{+1.01} \times 10^{44} $ & 132.0/70  & $\la 10^{-5}$  \\[0.1cm]
2 & J010415.8+402244 & 0.193 & 0.48 & $ <0.4 $ & $ 3.11_{-0.15}^{+0.42} $ & $ 3.61_{-0.18}^{+1.29} \times 10^{44} $ & 57.4/29  &$ 1.40 \times 10^{-4} $ \\[0.1cm]
3 & J061310.2+502628 & 0.019 & 1.58 & $ 2.31_{-0.60}^{+0.75} $ & $ 2.33_{-0.29}^{+0.35} $ & $ 1.06_{-0.10}^{+0.21} \times 10^{43} $ & 30.4/30  &$ 5.59 \times 10^{-1} $ \\[0.1cm]
4 & J122057.8+813600 &  & 0.58 & $ 0.39_{-0.28}^{+0.89} $ & $ 1.64_{-0.23}^{+0.53} $ &  & 31.6/35  &$ 8.31 \times 10^{-1} $ \\[0.1cm]
5 & J134954.6+432857 & 0.045 & 0.14 & $ 0.11_{-0.04}^{+0.45} $ & $ 2.15_{-0.11}^{+0.36} $ & $ 2.14_{-0.18}^{+0.29} \times 10^{43} $ & 17.6/31  &$ 7.46 \times 10^{-1} $ \\[0.1cm]
6 & J164147.7+700806 &  & 0.44 & $ 2.37_{-0.55}^{+0.73} $ & $ 4.63_{-0.46}^{+0.60} $ &  & 97.4/90  &$ 1.30 \times 10^{-4} $ \\[0.1cm]
7 & J161630.7+354230 & 0.028 & 0.11 & $ 0.20_{-0.12}^{+0.35} $ & $ 1.95_{-0.13}^{+0.28} $ & $ 8.50_{-0.90}^{+1.00} \times 10^{42} $ & 44.2/41  &$ 3.69 \times 10^{-1} $ \\[0.1cm]
8 & J025436.4+152545 &  & 0.76 & $ 0.53_{-0.38}^{+0.72} $ & $ 2.27_{-0.32}^{+0.53} $ &  & 25.5/29  &$ 9.61 \times 10^{-1} $ \\[0.1cm]
9 & J004730.3+154149 & 0.031 & 0.37 & $ 1.36_{-0.97}^{+1.48} $ & $ 2.40_{-0.52}^{+0.76} $ & $ 1.10_{-0.18}^{+0.72} \times 10^{43} $ & 27.7/28  &$ 8.07 \times 10^{-1} $ \\[0.1cm]
10 & J091935.9+753945 &  & 0.19 & $ 0.30_{-0.21}^{+0.51} $ & $ 2.46_{-0.23}^{+0.45} $ &  & 29.0/35  &$ 5.89 \times 10^{-1} $ \\[0.1cm]
11 & J024519.6-025627 &  & 0.31 & $ 1.51_{-0.76}^{+1.00} $ & $ 2.71_{-0.46}^{+0.58} $ &  & 22.1/31  &$ 1.85 \times 10^{-1} $ \\[0.1cm]
12 & J163323.5+471900 & 0.116 & 0.17 & $ <0.3 $ & $ 1.64_{-0.14}^{+0.24} $ & $ 6.98_{-1.12}^{+1.27} \times 10^{43} $ & 83.3/80  &$ 1.52 \times 10^{-2} $ \\[0.1cm]
13 & J174036.5+534624 &  & 0.29 & $ 0.67_{-0.34}^{+0.42} $ & $ 2.88_{-0.32}^{+0.37} $ &  & 96.2/104  &$ 4.54 \times 10^{-1} $ \\[0.1cm]
14 & J073203.1+680336 &  & 0.42 & $ <0.6 $ & $ 2.94_{-0.16}^{+0.69} $ &  & 67.8/67  &$ 4.37 \times 10^{-1} $ \\[0.1cm]
15 & J130600.3+330325 & 0.088 & 0.13 & $ 0.56_{-0.41}^{+1.39} $ & $ 2.37_{-0.33}^{+0.82} $ & $ 4.48_{-0.61}^{+3.15} \times 10^{43} $ & 23.7/27  &$ 4.25 \times 10^{-1} $ \\[0.1cm]
16 & J150311.8+111026 & 0.043 & 0.22 & $ 0.37_{-0.25}^{+0.76} $ & $ 1.93_{-0.25}^{+0.53} $ & $ 1.46_{-0.21}^{+0.32} \times 10^{43} $ & 48.5/40  &$ 5.53 \times 10^{-1} $ \\[0.1cm]
17 & J213946.2+113817 &  & 0.68 & $ 0.09_{-0.00}^{+1.07} $ & $ 1.04_{-0.09}^{+0.41} $ &  & 34.0/30  &$ 3.54 \times 10^{-1} $ \\[0.1cm]
18 & J160752.0+444125 & 0.718 & 0.09 & $ 0.08_{-0.01}^{+0.49} $ & $ 2.18_{-0.10}^{+0.42} $ & $ 6.40_{-0.29}^{+2.11} \times 10^{45} $ & 53.4/45  &$ 4.46 \times 10^{-1} $ \\[0.1cm]
19 & J135255.7+252900 & 0.064 & 0.11 & $ <0.7 $ & $ 1.63_{-0.12}^{+0.52} $ & $ 2.91_{-0.64}^{+0.62} \times 10^{43} $ & 36.2/33  &$ 5.83 \times 10^{-1} $ \\[0.1cm]
20 & J011739.4-025626 & 0.051 & 0.38 & $ <0.7 $ & $ 1.85_{-0.12}^{+0.45} $ & $ 2.04_{-0.31}^{+0.38} \times 10^{43} $ & 23.3/29  &$ 4.32 \times 10^{-1} $ \\[0.1cm]
21 & J175340.2+555128 &  & 0.37 & $ 0.47_{-0.33}^{+0.59} $ & $ 1.89_{-0.25}^{+0.40} $ &  & 83.3/103  &$ 8.70 \times 10^{-1} $ \\[0.1cm]
22 & J215717.0-263100 & 0.034 & 0.21 & $ <0.9 $ & $ 2.20_{-0.14}^{+0.70} $ & $ 7.40_{-1.02}^{+2.37} \times 10^{42} $ & 28.6/35  &$ 9.66 \times 10^{-1} $ \\[0.1cm]
23 & J145933.2+152742 & 0.074 & 0.19 & $ 0.60_{-0.43}^{+0.86} $ & $ 2.00_{-0.31}^{+0.54} $ & $ 4.48_{-0.59}^{+1.08} \times 10^{43} $ & 27.4/37  &$ 8.33 \times 10^{-1} $ \\[0.1cm]
24 & J000415.2+173454 &  & 0.28 & $ 3.90_{-1.41}^{+2.02} $ & $ 1.58_{-0.40}^{+0.54} $ &  & 26.5/29  &$ 1.40 \times 10^{-1} $ \\[0.1cm]
25 & J155926.1+521237 & 0.042 & 0.11 & $ 0.41_{-0.28}^{+0.50} $ & $ 2.23_{-0.25}^{+0.39} $ & $ 8.01_{-0.82}^{+1.34} \times 10^{42} $ & 68.9/62  &$ 5.90 \times 10^{-2} $ \\[0.1cm]
26 & J142408.5+210520 & 0.047 & 0.28 & $ 0.26_{-0.16}^{+1.01} $ & $ 1.80_{-0.22}^{+0.65} $ & $ 1.23_{-0.23}^{+0.31} \times 10^{43} $ & 32.9/35  &$ 9.51 \times 10^{-1} $ \\[0.1cm]
27 & J010812.6-114403 &  & 0.24 & $ 1.62_{-0.69}^{+0.86} $ & $ 5.48_{-0.68}^{+0.83} $ &  & 30.8/31  &$ 2.41 \times 10^{-2} $ \\[0.1cm]
28 & J135321.5+373055 & 0.107 & 0.14 & $ 0.72_{-0.51}^{+0.90} $ & $ 3.06_{-0.43}^{+0.72} $ & $ 7.81_{-1.83}^{+6.77} \times 10^{43} $ & 35.0/29  &$ 1.79 \times 10^{-1} $ \\[0.1cm]
29 & J150809.2+130031 & 0.086 & 0.24 & $ 0.64_{-0.45}^{+0.78} $ & $ 2.82_{-0.42}^{+0.65} $ & $ 4.21_{-0.81}^{+2.53} \times 10^{43} $ & 36.8/38  &$ 3.78 \times 10^{-1} $ \\[0.1cm]
30 & J155344.3+122233 & 0.035 & 0.35 & $ 0.74_{-0.45}^{+0.71} $ & $ 2.08_{-0.32}^{+0.45} $ & $ 1.05_{-0.12}^{+0.21} \times 10^{43} $ & 52.1/51  &$ 1.74 \times 10^{-2} $ \\[0.1cm]
31 & J153552.5+143104 & 0.020 & 0.35 & $ 0.48_{-0.32}^{+0.51} $ & $ 2.02_{-0.24}^{+0.36} $ & $ 5.55_{-0.61}^{+0.86} \times 10^{42} $ & 66.9/41  &$ 1.28 \times 10^{-2} $ \\[0.1cm]
32 & J124339.2+700515 &  & 0.14 & $ 0.40_{-0.29}^{+0.54} $ & $ 3.37_{-0.32}^{+0.53} $ &  & 26.5/30  &$ 8.34 \times 10^{-1} $ \\[0.1cm]
33 & J162012.8+400907 & 0.028 & 0.07 & $ 0.19_{-0.11}^{+0.21} $ & $ 2.46_{-0.12}^{+0.19} $ & $ 1.57_{-0.08}^{+0.13} \times 10^{43} $ & 56.9/46  &$ 9.00 \times 10^{-1} $ \\[0.1cm]
34 & J142607.6+340425 & 1.553 & 0.13 & $ 1.22_{-0.70}^{+1.01} $ & $ 3.09_{-0.51}^{+0.69} $ & $ 1.22_{-0.64}^{+2.93} \times 10^{47} $ & 24.4/31  &$ 2.43 \times 10^{-1} $ \\[0.1cm]
35 & J012633.6+313658 & 0.045 & 0.45 & $ 0.43_{-0.29}^{+0.52} $ & $ 2.33_{-0.26}^{+0.43} $ & $ 2.05_{-0.22}^{+0.40} \times 10^{43} $ & 44.9/31  &$ 1.24 \times 10^{-1} $ \\[0.1cm]
36 & J152410.3+485409 & 0.145 & 0.16 & $ <0.6 $ & $ 2.54_{-0.13}^{+0.50} $ & $ 7.56_{-0.42}^{+2.41} \times 10^{43} $ & 64.5/42  &$ 3.39 \times 10^{-3} $ \\[0.1cm]
37 & J121755.1+583936 & 0.023 & 0.15 & $ <0.3 $ & $ 2.53_{-0.16}^{+0.34} $ & $ 3.09_{-0.29}^{+0.55} \times 10^{42} $ & 35.3/28  &$ 1.48 \times 10^{-2} $ \\[0.1cm]
38 & J163016.2+584247 &  & 0.14 & $ 0.26_{-0.17}^{+0.29} $ & $ 2.02_{-0.16}^{+0.23} $ &  & 107.5/97  &$ 7.76 \times 10^{-2} $ \\[0.1cm]
39 & J035309.6+565431 &  & 4.48 & $ 11.13_{-1.97}^{+2.56} $ & $ 3.10_{-0.46}^{+0.60} $ &  & 32.1/28  &$ 2.24 \times 10^{-1} $ \\[0.1cm]
40 & J155829.3+271714 & 0.090 & 0.39 & $ 0.37_{-0.25}^{+0.37} $ & $ 1.89_{-0.19}^{+0.26} $ & $ 1.21_{-0.12}^{+0.15} \times 10^{44} $ & 243.2/212  &$ 9.96 \times 10^{-2} $ \\[0.1cm]
41 & J142352.1+245417 & 0.074 & 0.17 & $ <0.9 $ & $ 1.91_{-0.10}^{+0.52} $ & $ 4.37_{-0.54}^{+0.81} \times 10^{43} $ & 46.4/31  &$ 6.05 \times 10^{-1} $ \\[0.1cm]
42 & J142812.0+511115 & 0.129 & 0.11 & $ 0.21_{-0.13}^{+0.58} $ & $ 2.16_{-0.19}^{+0.47} $ & $ 1.01_{-0.11}^{+0.22} \times 10^{44} $ & 29.2/32  &$ 3.35 \times 10^{-1} $ \\[0.1cm]
43 & J160855.5+155200 & 0.115 & 0.30 & $ 0.38_{-0.27}^{+0.48} $ & $ 2.87_{-0.27}^{+0.41} $ & $ 1.61_{-0.22}^{+0.55} \times 10^{44} $ & 40.1/41  &$ 8.84 \times 10^{-1} $ \\[0.1cm]
44 & J130020.0+613919 & 0.052 & 0.15 & $ 0.35_{-0.25}^{+0.47} $ & $ 2.48_{-0.23}^{+0.38} $ & $ 2.96_{-0.27}^{+0.60} \times 10^{43} $ & 22.8/30  &$ 2.87 \times 10^{-1} $ \\[0.1cm]
45 & J131443.9+234826 & 2.060 & 0.10 & $ 0.78_{-0.50}^{+0.66} $ & $ 2.75_{-0.37}^{+0.46} $ & $ 3.28_{-1.34}^{+3.92} \times 10^{47} $ & 29.4/31  &$ 5.71 \times 10^{-1} $ \\[0.1cm]
46 & J181214.1+215305 & 0.018 & 0.86 & $ 2.19_{-0.89}^{+1.27} $ & $ 1.76_{-0.38}^{+0.50} $ & $ 6.51_{-0.87}^{+1.54} \times 10^{42} $ & 20.0/27  &$ 7.31 \times 10^{-1} $ \\[0.1cm]
47 & J021749.0+014449 & 1.715 & 0.34 & $ 0.49_{-0.33}^{+0.48} $ & $ 2.30_{-0.25}^{+0.35} $ & $ 1.31_{-0.29}^{+0.72} \times 10^{47} $ & 25.0/32  &$ 8.34 \times 10^{-1} $ \\[0.1cm]
48 & J181335.1+314418 & 0.117 & 0.44 & $ 1.04_{-0.34}^{+0.38} $ & $ 3.01_{-0.25}^{+0.27} $ & $ 7.39_{-1.17}^{+1.87} \times 10^{44} $ & 30.6/29  &$ 8.64 \times 10^{-3} $ \\[0.1cm]
49 & J022239.6+430208 & 0.444 & 0.82 & $ 1.28_{-0.35}^{+0.39} $ & $ 2.62_{-0.21}^{+0.23} $ & $ 2.48_{-0.37}^{+0.58} \times 10^{46} $ & 57.5/58  &$ 3.02 \times 10^{-1} $ \\[0.1cm]

%% file: main_en.bbl
\begin{thebibliography}{}
\makeatletter
\relax
\def\mn@urlcharsother{\let\do\@makeother \do\$\do\&\do\#\do\^\do\_\do\%\do\~}
\def\mn@doi{\begingroup\mn@urlcharsother \@ifnextchar [ {\mn@doi@}
  {\mn@doi@[]}}
\def\mn@doi@[#1]#2{\def\@tempa{#1}\ifx\@tempa\@empty \href
  {http://dx.doi.org/#2} {doi:#2}\else \href {http://dx.doi.org/#2} {#1}\fi
  \endgroup}
\def\mn@eprint#1#2{\mn@eprint@#1:#2::\@nil}
\def\mn@eprint@arXiv#1{\href {http://arxiv.org/abs/#1} {{\tt arXiv:#1}}}
\def\mn@eprint@dblp#1{\href {http://dblp.uni-trier.de/rec/bibtex/#1.xml}
  {dblp:#1}}
\def\mn@eprint@#1:#2:#3:#4\@nil{\def\@tempa {#1}\def\@tempb {#2}\def\@tempc
  {#3}\ifx \@tempc \@empty \let \@tempc \@tempb \let \@tempb \@tempa \fi \ifx
  \@tempb \@empty \def\@tempb {arXiv}\fi \@ifundefined
  {mn@eprint@\@tempb}{\@tempb:\@tempc}{\expandafter \expandafter \csname
  mn@eprint@\@tempb\endcsname \expandafter{\@tempc}}}

\bibitem[\protect\citeauthoryear{{Ahumada} et~al.,}{{Ahumada}
  et~al.}{2020}]{dr16}
{Ahumada} R.,  et~al., 2020, \mn@doi [\apjs] {10.3847/1538-4365/ab929e}, \href
  {https://ui.adsabs.harvard.edu/abs/2020ApJS..249....3A} {249, 3}

\bibitem[\protect\citeauthoryear{Anderson}{Anderson}{1962}]{anderson}
Anderson T.~W.,  1962, \mn@doi [The Annals of Mathematical Statistics]
  {10.1214/aoms/1177704477}, 33, 1148

\bibitem[\protect\citeauthoryear{{Antonucci}}{{Antonucci}}{1993}]{Antonucci}
{Antonucci} R.,  1993, \mn@doi [\araa] {10.1146/annurev.aa.31.090193.002353},
  \href {https://ui.adsabs.harvard.edu/abs/1993ARA&A..31..473A} {31, 473}

\bibitem[\protect\citeauthoryear{{Arnaud}}{{Arnaud}}{1996}]{Arnaud1996}
{Arnaud} K.~A.,  1996, in {Jacoby} G.~H.,  {Barnes} J.,  eds,  Astronomical
  Society of the Pacific Conference Series Vol. 101, Astronomical Data Analysis
  Software and Systems V. p.~17

\bibitem[\protect\citeauthoryear{{Assef} et~al.,}{{Assef}
  et~al.}{2013}]{Assef_2013}
{Assef} R.~J.,  et~al., 2013, \mn@doi [\apj] {10.1088/0004-637X/772/1/26},
  \href {https://ui.adsabs.harvard.edu/abs/2013ApJ...772...26A} {772, 26}

\bibitem[\protect\citeauthoryear{{Bellm} et~al.,}{{Bellm}
  et~al.}{2019}]{Bellm2019b}
{Bellm} E.~C.,  et~al., 2019, \mn@doi [\pasp] {10.1088/1538-3873/aaecbe}, \href
  {https://ui.adsabs.harvard.edu/abs/2019PASP..131a8002B} {131, 018002}

\bibitem[\protect\citeauthoryear{{Boisse} \& {Bergeron}}{{Boisse} \&
  {Bergeron}}{1988}]{boisse}
{Boisse} P.,  {Bergeron} J.,  1988, \aap, \href
  {https://ui.adsabs.harvard.edu/abs/1988A&A...192....1B} {192, 1}

\bibitem[\protect\citeauthoryear{{Brunner} et~al.,}{{Brunner}
  et~al.}{2022}]{brunner2022}
{Brunner} H.,  et~al., 2022, \mn@doi [\aap] {10.1051/0004-6361/202141266},
  \href {https://ui.adsabs.harvard.edu/abs/2022A&A...661A...1B} {661, A1}

\bibitem[\protect\citeauthoryear{{Cash}}{{Cash}}{1979}]{Cash1979}
{Cash} W.,  1979, \mn@doi [\apj] {10.1086/156922}, \href
  {https://ui.adsabs.harvard.edu/abs/1979ApJ...228..939C} {228, 939}

\bibitem[\protect\citeauthoryear{Cramér}{Cramér}{1928}]{cvm}
Cramér H.,  1928, \mn@doi [Scandinavian Actuarial Journal]
  {10.1080/03461238.1928.10416862}, 1928, 13

\bibitem[\protect\citeauthoryear{{Flesch}}{{Flesch}}{2015}]{milliquas2}
{Flesch} E.~W.,  2015, \mn@doi [\pasa] {10.1017/pasa.2015.10}, \href
  {https://ui.adsabs.harvard.edu/abs/2015PASA...32...10F} {32, e010}

\bibitem[\protect\citeauthoryear{{Flesch}}{{Flesch}}{2021}]{milliquas1}
{Flesch} E.~W.,  2021, arXiv e-prints, \href
  {https://ui.adsabs.harvard.edu/abs/2021arXiv210512985F} {p. arXiv:2105.12985}

\bibitem[\protect\citeauthoryear{{Gaia Collaboration} et~al.,}{{Gaia
  Collaboration} et~al.}{2021}]{GAIAEDR3}
{Gaia Collaboration} et~al., 2021, \aap, \href
  {https://ui.adsabs.harvard.edu/abs/2021A%26A...649A...1G} {649, A1}

\bibitem[\protect\citeauthoryear{{Gaia Collaboration} et~al.,}{{Gaia
  Collaboration} et~al.}{2022}]{gaia3}
{Gaia Collaboration} et~al., 2022, arXiv e-prints, \href
  {https://ui.adsabs.harvard.edu/abs/2022arXiv220605681G} {p. arXiv:2206.05681}

\bibitem[\protect\citeauthoryear{Geweke}{Geweke}{1992}]{geweke}
Geweke J.,  1992, Bayesian statistics, 4, 641

\bibitem[\protect\citeauthoryear{Gibson \& Brandt}{Gibson \&
  Brandt}{2012}]{gibson2012}
Gibson R.~R.,  Brandt W.,  2012, The Astrophysical Journal, 746, 54

\bibitem[\protect\citeauthoryear{{Giommi} et~al.,}{{Giommi}
  et~al.}{1991}]{exosat}
{Giommi} P.,  et~al., 1991, \mn@doi [\apj] {10.1086/170408}, \href
  {https://ui.adsabs.harvard.edu/abs/1991ApJ...378...77G} {378, 77}

\bibitem[\protect\citeauthoryear{{Goodman} \& {Weare}}{{Goodman} \&
  {Weare}}{2010}]{emcee}
{Goodman} J.,  {Weare} J.,  2010, \mn@doi [Communications in Applied
  Mathematics and Computational Science] {10.2140/camcos.2010.5.65}, \href
  {https://ui.adsabs.harvard.edu/abs/2010CAMCS...5...65G} {5, 65}

\bibitem[\protect\citeauthoryear{{Graham} et~al.,}{{Graham}
  et~al.}{2019}]{Graham2019}
{Graham} M.~J.,  et~al., 2019, \mn@doi [\pasp] {10.1088/1538-3873/ab006c},
  \href {https://ui.adsabs.harvard.edu/abs/2019PASP..131g8001G} {131, 078001}

\bibitem[\protect\citeauthoryear{{HI4PI Collaboration} et~al.,}{{HI4PI
  Collaboration} et~al.}{2016}]{HI4PI_collab}
{HI4PI Collaboration} et~al., 2016, \mn@doi [\aap]
  {10.1051/0004-6361/201629178}, \href
  {https://ui.adsabs.harvard.edu/abs/2016A&A...594A.116H} {594, A116}

\bibitem[\protect\citeauthoryear{Hinkley}{Hinkley}{1969}]{hinkley1969ratio}
Hinkley D.~V.,  1969, Biometrika, 56, 635

\bibitem[\protect\citeauthoryear{{Huo} et~al.,}{{Huo} et~al.}{2013}]{lamost}
{Huo} Z.-Y.,  et~al., 2013, \mn@doi [\aj] {10.1088/0004-6256/145/6/159}, \href
  {https://ui.adsabs.harvard.edu/abs/2013AJ....145..159H} {145, 159}

\bibitem[\protect\citeauthoryear{{Jones} et~al.,}{{Jones} et~al.}{2009}]{6df}
{Jones} D.~H.,  et~al., 2009, \mn@doi [\mnras]
  {10.1111/j.1365-2966.2009.15338.x}, \href
  {https://ui.adsabs.harvard.edu/abs/2009MNRAS.399..683J} {399, 683}

\bibitem[\protect\citeauthoryear{Lanzuisi et~al.,}{Lanzuisi
  et~al.}{2014}]{lanzuisi2014}
Lanzuisi G.,  et~al., 2014, The Astrophysical Journal, 781, 105

\bibitem[\protect\citeauthoryear{{Lawrence}, {Watson}, {Pounds}  \&
  {Elvis}}{{Lawrence} et~al.}{1987}]{lawrence}
{Lawrence} A.,  {Watson} M.~G.,  {Pounds} K.~A.,   {Elvis} M.,  1987, \mn@doi
  [\nat] {10.1038/325694a0}, \href
  {https://ui.adsabs.harvard.edu/abs/1987Natur.325..694L} {325, 694}

\bibitem[\protect\citeauthoryear{{Liu}, {Liu}, {Dong}, {Zhou}, {Wang}, {Lu}  \&
  {Yuan}}{{Liu} et~al.}{2019}]{lozagn}
{Liu} H.-Y.,  {Liu} W.-J.,  {Dong} X.-B.,  {Zhou} H.,  {Wang} T.,  {Lu} H.,
  {Yuan} W.,  2019, \mn@doi [\apjs] {10.3847/1538-4365/ab298b}, \href
  {https://ui.adsabs.harvard.edu/abs/2019ApJS..243...21L} {243, 21}

\bibitem[\protect\citeauthoryear{{Lyke} et~al.,}{{Lyke} et~al.}{2020}]{dr16q}
{Lyke} B.~W.,  et~al., 2020, \mn@doi [\apjs] {10.3847/1538-4365/aba623}, \href
  {https://ui.adsabs.harvard.edu/abs/2020ApJS..250....8L} {250, 8}

\bibitem[\protect\citeauthoryear{{Markowitz} \& {Edelson}}{{Markowitz} \&
  {Edelson}}{2004}]{markowitz2004}
{Markowitz} A.,  {Edelson} R.,  2004, \mn@doi [\apj] {10.1086/425559}, \href
  {https://ui.adsabs.harvard.edu/abs/2004ApJ...617..939M} {617, 939}

\bibitem[\protect\citeauthoryear{{Masci} et~al.,}{{Masci}
  et~al.}{2019}]{Masci2019}
{Masci} F.~J.,  et~al., 2019, \mn@doi [\pasp] {10.1088/1538-3873/aae8ac}, \href
  {http://adsabs.harvard.edu/abs/2019PASP..131a8003M} {131, 018003}

\bibitem[\protect\citeauthoryear{{Massaro}, {Giommi}, {Leto}, {Marchegiani},
  {Maselli}, {Perri}, {Piranomonte}  \& {Sclavi}}{{Massaro}
  et~al.}{2009}]{bzcat}
{Massaro} E.,  {Giommi} P.,  {Leto} C.,  {Marchegiani} P.,  {Maselli} A.,
  {Perri} M.,  {Piranomonte} S.,   {Sclavi} S.,  2009, \mn@doi [\aap]
  {10.1051/0004-6361:200810161}, \href
  {https://ui.adsabs.harvard.edu/abs/2009A&A...495..691M} {495, 691}

\bibitem[\protect\citeauthoryear{Matt, Guainazzi  \& Maiolino}{Matt
  et~al.}{2003}]{matt}
Matt G.,  Guainazzi M.,   Maiolino R.,  2003, \mn@doi [Monthly Notices of the
  Royal Astronomical Society] {10.1046/j.1365-8711.2003.06539.x}, 342, 422

\bibitem[\protect\citeauthoryear{{McHardy} \& {Czerny}}{{McHardy} \&
  {Czerny}}{1987}]{mchardy}
{McHardy} I.,  {Czerny} B.,  1987, \mn@doi [\nat] {10.1038/325696a0}, \href
  {https://ui.adsabs.harvard.edu/abs/1987Natur.325..696M} {325, 696}

\bibitem[\protect\citeauthoryear{Middei, Vagnetti, Bianchi, La~Franca, Paolillo
   \& Ursini}{Middei et~al.}{2017}]{middei2017}
Middei R.,  Vagnetti F.,  Bianchi S.,  La~Franca F.,  Paolillo M.,   Ursini F.,
   2017, Astronomy \& Astrophysics, 599, A82

\bibitem[\protect\citeauthoryear{{Osterbrock}}{{Osterbrock}}{1977}]{osterbrock}
{Osterbrock} D.~E.,  1977, \mn@doi [\apj] {10.1086/155407}, \href
  {https://ui.adsabs.harvard.edu/abs/1977ApJ...215..733O} {215, 733}

\bibitem[\protect\citeauthoryear{{Pavlinsky} et~al.,}{{Pavlinsky}
  et~al.}{2021}]{2021AA...650A..42P}
{Pavlinsky} M.,  et~al., 2021, \mn@doi [\aap] {10.1051/0004-6361/202040265},
  \href {https://ui.adsabs.harvard.edu/abs/2021A&A...650A..42P} {650, A42}

\bibitem[\protect\citeauthoryear{{Peterson}}{{Peterson}}{2001}]{peterson2001}
{Peterson} B.~M.,  2001, in {Aretxaga} I.,  {Kunth} D.,   {M{\'u}jica} R.,
  eds, Advanced Lectures on the Starburst-AGN. p.~3 (\mn@eprint {arXiv}
  {astro-ph/0109495}), \mn@doi{10.1142/9789812811318_0002}

\bibitem[\protect\citeauthoryear{{Predehl} et~al.,}{{Predehl}
  et~al.}{2021}]{2021AA...647A...1P}
{Predehl} P.,  et~al., 2021, \mn@doi [\aap] {10.1051/0004-6361/202039313},
  \href {https://ui.adsabs.harvard.edu/abs/2021A&A...647A...1P} {647, A1}

\bibitem[\protect\citeauthoryear{Puccetti et~al.,}{Puccetti
  et~al.}{2014}]{Puccetti_2014}
Puccetti S.,  et~al., 2014, \mn@doi [The Astrophysical Journal]
  {10.1088/0004-637X/793/1/26}, 793, 26

\bibitem[\protect\citeauthoryear{Ricci et~al.,}{Ricci
  et~al.}{2016}]{Ricci_2016}
Ricci C.,  et~al., 2016, \mn@doi [The Astrophysical Journal]
  {10.3847/0004-637X/820/1/5}, 820, 5

\bibitem[\protect\citeauthoryear{Ricci et~al.,}{Ricci
  et~al.}{2020}]{Ricci_2020}
Ricci C.,  et~al., 2020, \mn@doi [The Astrophysical Journal Letters]
  {10.3847/2041-8213/ab91a1}, 898, L1

\bibitem[\protect\citeauthoryear{{Sazonov} et~al.,}{{Sazonov}
  et~al.}{2021}]{sazonov2021}
{Sazonov} S.,  et~al., 2021, \mn@doi [\mnras] {10.1093/mnras/stab2843}, \href
  {https://ui.adsabs.harvard.edu/abs/2021MNRAS.508.3820S} {508, 3820}

\bibitem[\protect\citeauthoryear{{Shakura} \& {Sunyaev}}{{Shakura} \&
  {Sunyaev}}{1973}]{shakura}
{Shakura} N.~I.,  {Sunyaev} R.~A.,  1973, \aap, \href
  {https://ui.adsabs.harvard.edu/abs/1973A&A....24..337S} {24, 337}

\bibitem[\protect\citeauthoryear{{Shakura} \& {Sunyaev}}{{Shakura} \&
  {Sunyaev}}{1976}]{Shakura76}
{Shakura} N.~I.,  {Sunyaev} R.~A.,  1976, \mn@doi [\mnras]
  {10.1093/mnras/175.3.613}, \href
  {https://ui.adsabs.harvard.edu/abs/1976MNRAS.175..613S} {175, 613}

\bibitem[\protect\citeauthoryear{{Shemmer} et~al.,}{{Shemmer}
  et~al.}{2014}]{shemmer2014}
{Shemmer} O.,  et~al., 2014, \mn@doi [\apj] {10.1088/0004-637X/783/2/116},
  \href {https://ui.adsabs.harvard.edu/abs/2014ApJ...783..116S} {783, 116}

\bibitem[\protect\citeauthoryear{Shemmer, Brandt, Paolillo, Kaspi, Vignali,
  Lira  \& Schneider}{Shemmer et~al.}{2017}]{shemmer2017}
Shemmer O.,  Brandt W.~N.,  Paolillo M.,  Kaspi S.,  Vignali C.,  Lira P.,
  Schneider D.~P.,  2017, The Astrophysical Journal, 848, 46

\bibitem[\protect\citeauthoryear{{Smith} et~al.,}{{Smith}
  et~al.}{2020}]{Smith2020}
{Smith} K.~W.,  et~al., 2020, \mn@doi [\pasp] {10.1088/1538-3873/ab936e}, \href
  {https://ui.adsabs.harvard.edu/abs/2020PASP..132h5002S} {132, 085002}

\bibitem[\protect\citeauthoryear{{Stern} et~al.,}{{Stern} et~al.}{2018}]{Stern}
{Stern} D.,  et~al., 2018, \mn@doi [\apj] {10.3847/1538-4357/aac726}, \href
  {https://ui.adsabs.harvard.edu/abs/2018ApJ...864...27S} {864, 27}

\bibitem[\protect\citeauthoryear{{Sunyaev} et~al.,}{{Sunyaev}
  et~al.}{2021}]{2021AA...656A.132S}
{Sunyaev} R.,  et~al., 2021, \mn@doi [\aap] {10.1051/0004-6361/202141179},
  \href {https://ui.adsabs.harvard.edu/abs/2021A&A...656A.132S} {656, A132}

\bibitem[\protect\citeauthoryear{{The Fermi-LAT collaboration} et~al.,}{{The
  Fermi-LAT collaboration} et~al.}{2022a}]{fermi}
{The Fermi-LAT collaboration} et~al., 2022a, arXiv e-prints, \href
  {https://ui.adsabs.harvard.edu/abs/2022arXiv220912070T} {p. arXiv:2209.12070}

\bibitem[\protect\citeauthoryear{{The Fermi-LAT collaboration} et~al.,}{{The
  Fermi-LAT collaboration} et~al.}{2022b}]{fermi_dr3}
{The Fermi-LAT collaboration} et~al., 2022b, arXiv e-prints, \href
  {https://ui.adsabs.harvard.edu/abs/2022arXiv220912070T} {p. arXiv:2209.12070}

\bibitem[\protect\citeauthoryear{Timlin~III, Brandt, Zhu, Liu, Luo  \&
  Ni}{Timlin~III et~al.}{2020}]{timlin2020}
Timlin~III J.~D.,  Brandt W.~N.,  Zhu S.,  Liu H.,  Luo B.,   Ni Q.,  2020,
  Monthly Notices of the Royal Astronomical Society, 498, 4033

\bibitem[\protect\citeauthoryear{{Tonry} et~al.,}{{Tonry}
  et~al.}{2018}]{Tonry2018}
{Tonry} J.~L.,  et~al., 2018, \mn@doi [\pasp] {10.1088/1538-3873/aabadf}, \href
  {https://ui.adsabs.harvard.edu/abs/2018PASP..130f4505T} {130, 064505}

\bibitem[\protect\citeauthoryear{Uttley, McHardy  \& Papadakis}{Uttley
  et~al.}{2002}]{uttley}
Uttley P.,  McHardy I.~M.,   Papadakis I.~E.,  2002, \mn@doi [Monthly Notices
  of the Royal Astronomical Society] {10.1046/j.1365-8711.2002.05298.x}, 332,
  231

\bibitem[\protect\citeauthoryear{{Vagnetti}, {Turriziani}  \&
  {Trevese}}{{Vagnetti} et~al.}{2011}]{vagnetti2011}
{Vagnetti} F.,  {Turriziani} S.,   {Trevese} D.,  2011, \mn@doi [\aap]
  {10.1051/0004-6361/201118072}, \href
  {https://ui.adsabs.harvard.edu/abs/2011A&A...536A..84V} {536, A84}

\bibitem[\protect\citeauthoryear{Vagnetti, Middei, Antonucci, Paolillo  \&
  Serafinelli}{Vagnetti et~al.}{2016}]{vagnetti2016}
Vagnetti F.,  Middei R.,  Antonucci M.,  Paolillo M.,   Serafinelli R.,  2016,
  Astronomy \& Astrophysics, 593, A55

\bibitem[\protect\citeauthoryear{{Veron}, {Goncalves}  \&
  {Veron-Cetty}}{{Veron} et~al.}{1997}]{veron}
{Veron} P.,  {Goncalves} A.~C.,   {Veron-Cetty} M.~P.,  1997, \aap, \href
  {https://ui.adsabs.harvard.edu/abs/1997A&A...319...52V} {319, 52}

\bibitem[\protect\citeauthoryear{{Wake} et~al.,}{{Wake}
  et~al.}{2017}]{sdss-manga}
{Wake} D.~A.,  et~al., 2017, \mn@doi [\aj] {10.3847/1538-3881/aa7ecc}, \href
  {https://ui.adsabs.harvard.edu/abs/2017AJ....154...86W} {154, 86}

\bibitem[\protect\citeauthoryear{White et~al.,}{White et~al.}{2000}]{first}
White R.~L.,  et~al., 2000, \mn@doi [The Astrophysical Journal Supplement
  Series] {10.1086/313300}, 126, 133

\bibitem[\protect\citeauthoryear{{Wilms}, {Allen}  \& {McCray}}{{Wilms}
  et~al.}{2000}]{wilms2000}
{Wilms} J.,  {Allen} A.,   {McCray} R.,  2000, \mn@doi [\apj] {10.1086/317016},
  \href {https://ui.adsabs.harvard.edu/abs/2000ApJ...542..914W} {542, 914}

\bibitem[\protect\citeauthoryear{{Wright} et~al.,}{{Wright}
  et~al.}{2010}]{Wright_2010}
{Wright} E.~L.,  et~al., 2010, \mn@doi [\aj] {10.1088/0004-6256/140/6/1868},
  \href {https://ui.adsabs.harvard.edu/abs/2010AJ....140.1868W} {140, 1868}

\bibitem[\protect\citeauthoryear{{Yao} et~al.,}{{Yao} et~al.}{2019}]{lamost_q5}
{Yao} S.,  et~al., 2019, \mn@doi [\apjs] {10.3847/1538-4365/aaef88}, \href
  {https://ui.adsabs.harvard.edu/abs/2019ApJS..240....6Y} {240, 6}

\bibitem[\protect\citeauthoryear{{Zaw}, {Chen}  \& {Farrar}}{{Zaw}
  et~al.}{2019}]{twomass}
{Zaw} I.,  {Chen} Y.-P.,   {Farrar} G.~R.,  2019, \mn@doi [\apj]
  {10.3847/1538-4357/aaffaf}, \href
  {https://ui.adsabs.harvard.edu/abs/2019ApJ...872..134Z} {872, 134}

\bibitem[\protect\citeauthoryear{Álvarez Crespo et~al.,}{Álvarez Crespo
  et~al.}{2016}]{crespo_2016}
Álvarez Crespo N.,  et~al., 2016, \mn@doi [The Astronomical Journal]
  {10.3847/0004-6256/151/4/95}, 151, 95

\makeatother
\end{thebibliography}
